
\input phyzzx.tex

 \overfullrule=0pt
 \Pubnum={\vbox { \hbox{CERN-TH.6242/91}\hbox{UGVA-PHY-09-745/91}}}
 \pubnum={CERN-TH.6242/91}
 \date={September, 1991}
 \pubtype={}
 \titlepage


\def\Zot{{ Z_{12} }}
\def\Zoth{{ Z_{13} }}
\def\Ztth{{ Z_{23} }}
\def\Tot{{ \theta_{12} }}
\def\alnot{{ \alpha_0 }}
\def\alm{{ \alpha_- }}
\def\alp{{ \alpha_+ }}
\def\almm{{ \alpha_{m',m} }}
\def\mod{{ \rm \;mod\; }}
\def\half{{{1 \over 2}}}
\def\wtilde#1{{ \widetilde#1}}
\def\jkm{{ J_k^{(m)} }}
\def\ikm{{ I_k^{(m)} }}
\def\nkm{{ N_k^{(m)} }}
\def\jlknm{{ J_{lk}^{(nm)} }}
\def\ilknm{{ I_{lk}^{(nm)} }}
\def\nlknm{{ N_{lk}^{(nm)} }}
\def\tjkm{{ \wtilde{J}_k^{(m)} }}
\def\tikm{{ \wtilde{I}_k^{(m)} }}
\def\zhf{{ z^{1/2} }}
\def\abcr{{ a,b,c,\rho }}
\def\iti{{ \it i}}
\def\itii{{ \it ii}}
\def\itiii{{ \it iii}}
\def\itiv{{ \it iv}}
\def\CFkm{{ {\cal F}_k^{(m)} }}
\def\CFlknm{{ {\cal F}_{lk}^{(nm)} }}
\def\skm{{ S_k^{(m)} }}
\def\slknm{{ S_{lk}^{(nm)} }}
\def\cnsp{{ C_{NS}^P }}
\def\csnp{{ C_{SN}^P }}
\def\rom{{ {\rho-1 \over 2} }}
\def\rop{{ {\rho+1 \over 2} }}
\def\ropm{{ {\rho'-1 \over 2} }}
\def\ropp{{ {\rho'+1 \over 2} }}
\def\siro{{ s(i\rom) }}

\def\albr{{ \alpha,\beta,\rho }}
\def\gomro{{ \Gamma({1-\rho \over 2}) }}
\def\gomrop{{ \Gamma({1-\rho' \over 2}) }}
\def\jnm{{ J_{nm} }}
\def\tjs{{ \wtilde{J}^\Sigma }}
\def\js{{ J^\Sigma }}
\def\tjnms{{ \wtilde{J}^\Sigma_{nm} }}
\def\jnms{{ \js_{nm} }}
\def\J#1 #2 #3 #4{{ \Big( \matrix{\scriptstyle #1 &\scriptstyle #2 \cr
  \scriptstyle #3 &\scriptstyle #4\cr}\Big) }}
\def\JJ#1 #2 #3 #4 #5 #6{{ \Big( \matrix{\scriptstyle #1 &\scriptstyle #2 &
  \scriptstyle #3\cr \scriptstyle #4 &\scriptstyle #5 &\scriptstyle #6\cr}
  \Big) }}
\def\oeps{{ 1_\epsilon }}
\def\dsnp{{ D_{SN}^P }}
\def\sltwo{SL(2|1)}



\def\(#1){(\call{#1})}
\def\call#1{{#1}}
\def\taghead#1{}

\catcode`@=11
\newcount\tagnumber\tagnumber=0
\immediate\newwrite\eqnfile
\newif\if@qnfile\@qnfilefalse
\def\write@qn#1{}
\def\writenew@qn#1{}
\def\w@rnwrite#1{\write@qn{#1}\message{#1}}
\def\@rrwrite#1{\write@qn{#1}\errmessage{#1}}

\def\taghead#1{\gdef\t@ghead{#1}\global\tagnumber=0}
\def\t@ghead{}

\expandafter\def\csname @qnnum -3\endcsname
  {{\t@ghead\advance\tagnumber by -3\relax\number\tagnumber}}
\expandafter\def\csname @qnnum -2\endcsname
  {{\t@ghead\advance\tagnumber by -2\relax\number\tagnumber}}
\expandafter\def\csname @qnnum -1\endcsname
  {{\t@ghead\advance\tagnumber by -1\relax\number\tagnumber}}
\expandafter\def\csname @qnnum0\endcsname
  {\t@ghead\number\tagnumber}
\expandafter\def\csname @qnnum+1\endcsname
  {{\t@ghead\advance\tagnumber by 1\relax\number\tagnumber}}
\expandafter\def\csname @qnnum+2\endcsname
  {{\t@ghead\advance\tagnumber by 2\relax\number\tagnumber}}
\expandafter\def\csname @qnnum+3\endcsname
  {{\t@ghead\advance\tagnumber by 3\relax\number\tagnumber}}

\def\equationfile{%
  \@qnfiletrue\immediate\openout\eqnfile=\jobname.eqn%
  \def\write@qn##1{\if@qnfile\immediate\write\eqnfile{##1}\fi}
  \def\writenew@qn##1{\if@qnfile\immediate\write\eqnfile
     {\noexpand\tag{##1} = (\t@ghead\number\tagnumber)}\fi}
     }
\def\callall#1{\xdef#1##1{#1{\noexpand\call{##1}}}}
\def\call#1{\each@rg\callr@nge{#1}}

\def\each@rg#1#2{{\let\thecsname=#1\expandafter\first@rg#2,\end,}}
\def\first@rg#1,{\thecsname{#1}\apply@rg}
\def\apply@rg#1,{\ifx\end#1\let\next=\relax%
\else,\thecsname{#1}\let\next=\apply@rg\fi\next}

\def\callr@nge#1{\calldor@nge#1-\end -}
\def\callr@ngeat#1\end -{#1}
\def\calldor@nge#1-#2-{\ifx\end#2\@qneatspace#1 %
  \else\calll@@p{#1}{#2}\callr@ngeat\fi}
\def\calll@@p#1#2{\ifnum#1>#2{\@rrwrite{Equation range #1-#2\space is bad.}
\errhelp{If you call a series of equations by the notation M-N, then M and
N must be integers, and N must be greater than or equal to M.}}\else%
 {\count0=#1\count1=#2\advance\count1 by1\relax\expandafter\@qncall
  \the\count0,%
  \loop\advance\count0 by1\relax%
    \ifnum\count0<\count1,\expandafter\@qncall\the\count0,%
  \repeat}\fi}

\def\@qneatspace#1#2 {\@qncall#1#2,}
\def\@qncall#1,{\ifunc@lled{#1}{\def\next{#1}\ifx\next\empty\else
  \w@rnwrite{Equation number \noexpand\(>>#1<<) has not been defined yet.}
  >>#1<<\fi}\else\csname @qnnum#1\endcsname\fi}

\let\eqnono=\eqno
\def\eqno(#1){\tag#1}
\def\tag#1$${\eqnono(\displayt@g#1 )$$}

\def\aligntag#1\endaligntag
  $${\gdef\tag##1\\{&(##1 )\cr}\eqalignno{#1\\}$$
  \gdef\tag##1$${\eqnono(\displayt@g##1 )$$}}

\def\eqalignno#1{\displ@y \tabskip\centering
  \halign to\displaywidth{\hfil$\displaystyle{##}$\tabskip\z@skip
    &$\displaystyle{{}##}$\hfil\tabskip\centering
    &\llap{$\displayt@gpar##$}\tabskip\z@skip\crcr
    #1\crcr}}

\def\displayt@gpar(#1){(\displayt@g#1 )}
\def\displayt@g#1 {\rm\ifunc@lled{#1}\global\advance\tagnumber by1
        {\def\next{#1}\ifx\next\empty\else\expandafter
        \xdef\csname @qnnum#1\endcsname{\t@ghead\number\tagnumber}\fi}%
  \writenew@qn{#1}\t@ghead\number\tagnumber\else
        {\edef\next{\t@ghead\number\tagnumber}%
        \expandafter\ifx\csname @qnnum#1\endcsname\next\else
        \w@rnwrite{Equation \noexpand\tag{#1} is a duplicate number.}\fi}%
  \csname @qnnum#1\endcsname\fi}

\def\ifunc@lled#1{\expandafter\ifx\csname @qnnum#1\endcsname\relax}
\let\@qnend=\end\gdef\end{\if@qnfile
\immediate\write16{Equation numbers written on []\jobname.EQN.}\fi\@qnend}
\catcode`@=12


\title{STRUCTURE CONSTANTS IN THE N=1 SUPEROPERATOR ALGEBRA}
\author{L. Alvarez-Gaum\'e}
\address{Theory Division, CERN\break CH-1211 Geneva 23, Switzerland}

\author{Ph. Zaugg$^*$}\foot{Partially supported by the Swiss National
 Science Foundation.}
\address{D\'epartement de Physique Th\'eorique\break
Universit\'e de Gen\`eve\break CH-1211 Gen\`eve 4, Switzerland}

\abstract{Using the Coulomb gas formulation of N=1 Superconformal
Field Theories, we extend the arguments of Dotsenko and Fateev
for the bosonic case to evaluate the structure constants of N=1
minimal Superconformal Algebras in the Neveu-Schwarz sector.}

\def\rdotfatone{ V.S. Dotsenko and V.A. Fateev: Nucl. Phys.{\bf B240}
 (1984) 312.}

\def\rdotfattwo{ V.S. Dotsenko and V.A. Fateev:  Nucl. Phys.{\bf B251}
 (1985) 691.}

\def\rdotfatthree{ V.S. Dotsenko and V.A. Fateev: Phys. Lett {\bf 154B}
 (1985) 291.}

\def\rbpz{A.A. Belavin, A.M. Polyakov and A.B. Zamolodchikov:
 Nucl. Phys.{\bf B241} (1984) 333.}

\def\rfqsone{D. Friedan, Z. Qiu and S. Shenker: In {\it Vertex Operators in
Mathematical Physics}. J. Lepowsky ed. Springer Verlag, 1984.}

\def\rfqstwo{D. Friedan, Z. Qiu and S. Shenker: Phys. Rev. Lett. {\bf 51}
 (1984)
1575.}

\def\rfqsthree{D. Friedan, Z. Qiu and S. Shenker: Phys. Lett. {\bf 151B}
 (1985) 37.}

\def\rfriedan{D. Friedan: {\it Notes on String Theory and
 Two Dimensional Conformal
Field Theory}.  In Proceedings of the Santa Barbara Workshop.
  M.B. Green and D.J. Gross eds.
World Scientific 1985.}

\def\rfms{D. Friedan, E. Martinec and S. Shenker: Nucl. Phys.
 {\bf B271} (1986) 93.}

\def\rbkt{M. Bershadsky, V. Knizhnik and A. Teitelman: Phys. Lett.
 {\bf 151B} (1985) 31.}

\def\rams{J. Atick, G. Moore and A. Sen:  Nucl. Phys.
 {\bf B308} (1988) 1.}

\def\reich{H. Eichenherr: Phys. Lett. {\bf 151B} (1985) 26.}

\def\rqiu{Z. Qiu: Nucl. Phys. {\bf B270} (1986) 205.}

\def\rmussardo{G. Mussardo, G. Sotkov and H. Stanishkov:
  Phys. Lett. {\bf 195B} (1987) 397;
Nucl. Phys. {\bf B305} (1988) 69.}

\def\rince{E.L. Ince {\it Ordinary Differential Equations}.
  Dover 1927.}


\endpage
 \pagenumber=1
\chapter{INTRODUCTION}

In this paper we compute the structure constants of
 the operator algebra for
some Superconformal Field Theories in the Neveu-Schwarz
 sector (NS).  For the
minimal Conformally Invariant Theories
\REF\bpz{\rbpz}[\bpz]
and for the unitary subseries,
\REF\fqstwo{\rfqstwo}[\fqstwo] the structure constants of
 the operator algebra
were computed in general using the Coulomb gas representation
 for the models
in a set of classic papers by Dotsenko and Fateev
\REF\dotfatone{\rdotfatone}
\REF\dotfattwo{\rdotfattwo}
\REF\dotfatthree{\rdotfatthree}
[\dotfatone,\dotfattwo,\dotfatthree]. The $N=1$ Superconformal
 Field Theories
\REF\fqsone{\rfqsone}
\REF\fqsthree{\rfqsthree}
[\fqsone,\fqstwo,\fqsthree] were found soon after the discovery of the
minimal models [\bpz], and some of their properties were analyzed in
\REF\eich{\reich}
\REF\bkt{\rbkt}
\REF\qiu{\rqiu}
[\eich,\bkt,\qiu].  In particular, in [\bkt] the Coulomb gas description
in [\dotfatone] was extended to the superconformal case.  This construction
was used to determine the fusion rules and some of the general properties of
four-point correlators
\REF\mussardo{\rmussardo}[\mussardo].  However, the full determination
of the structure constants of the operator algebra for the minimal
superconformal theories in analogy with the analysis of
[\dotfatone,\dotfattwo,\dotfatthree] is not available in the literature.

We extend the contour manipulation techniques of the work of
Dotsenko and Fateev to the case of supercontours
and supercontour integral representations
of superconformal blocks. We follow their methodology
 closely, although the
superconformal case presents some peculiarities of its own.
  There are several
ways to test the accuracy of our results.  The first one consists of
verifying that the structure constants have zeroes exactly where indicated
by the fusion rules.  We can also check that our results agree with those
of the tricritical Ising model, where the structure constants can be read
off directly from [\dotfattwo,\dotfatthree] or [\qiu].  The third and less
trivial check of the consistency of our method is presented in the appendix.
It
will be argued in the text that in the computation of the
 structure constants
we need to know two things.  First we need to know a set of
monodromy matrices expressing the behavior of the conformal blocks under
braiding and second we need to know explicitly a number of normalization
superintegrals.  There are two independent ways of computing the monodromy
matrices.  The first one, as we will discuss in detail, is to take them
from [\dotfattwo] after some appropriate changes are made, and the second is
to evaluate them directly in terms of the normalization integrals.
  These two
methods are independent and give the same result thus providing a good
verification of our results and methods.  This paper is very
 technical and for
the reader interested only in the results, we have summarized
 the main results
at the end of each section.

The structure of this paper is as follows.  In section two
 we collect a number of
useful formulae in the theory of superconformally invariant
 field theories.  We
have followed the presentations in
\REF\friedan{\rfriedan}
\REF\fms{\rfms}
[\friedan,\fms] and in [\bkt] for the Coulomb gas formulation
 of the Superconformal
minimal models.  We write the Coulomb gas representation of the
 chiral blocks
which need to be computed and at the end of the section we have a short
discussion on the issue of open versus closed supercontours in the
representation of  superconformal blocks and in the solution of
super-differential equations.  Here, and perhaps unexpectedly, one finds a
phenomenon familiar in Superstring perturbation theory with
 regard to integration
ambiguities in supermanifolds with boundaries
 (see \REF\ams{\rams} for details
and references) although here the problem is not nearly as severe.  We also
collect a number of useful formula about $SL(2|1)$ and
 its invariants which are
necessary in writing  three- and four-point correlators.

In section three we
extend to the  superconformal case the general analysis
 of conformal blocks that
was  carried out for the conformal case in [\bpz] and [\dotfattwo].
  The general
structure of the operator product expansion (OPE) of two superfield is
presented, we define the monodromy matrices and the monodromy invariants used
in
building physical correlation functions.  We write explicitly
 the holomorphic
superconformal blocks to be computed in later sections,
 we determine carefully
the  generalization of the conformal block normalization
 conditions in [\bpz]
and itemize all the possible case and how to determine
 the general form of the
monodromy invariant metric allowing us to put together the holomorphic
and anti-holomorphic blocks.  We do this for the thermal subalgebras and the
general algebra of NS superfields.  We determine here the type of integrals
we need to compute in order to explicitly evaluate the structure constants
of the superoperator algebra.  This section therefore
 outlines the computation
to follow.

In section four we compute the normalization integrals for the thermal
series.  We have to separate two substantially different
 cases depending on whether
the number of screening charges is even or odd.  The odd case
 is more difficult than
the even case because two types of integrals are necessary.  We
find the generalization of the recursion relations and
 functional relations for
them which generalize the work of [\dotfattwo], relate the even to the odd
integrals by some specific limiting procedures and we
 are able finally to write
the  explicit form of the integrals in terms of rather long products of
$\Gamma$-functions. The most difficult part of the
 computation appears in the
determination of a  set of integers appearing in the arguments of the
$\Gamma$-functions.

In section five we compute the normalization constants in
the general case.  Again the most difficult part is the determination of some
integers in the arguments of the $\Gamma$-functions,
 however, here we can use
the results of the thermal series to determine them.  In section six we
compute the non-symmetric structure constant of the superoperator algebra.
These are the ones we can read off  directly from the
 formulae in section three.
At the end of this section we collect a number of
 useful integrals analogous to
those appearing in appendix B of [\dotfattwo].  In
 section seven we extend the
arguments in [\dotfatthree] to compute the symmetric (physical) structure
constants.  Finally, in the appendix we provide a
 way to compute the monodromy
matrices different from the one used in the text.
  We believe that the methods
presented in this paper can be extended to compute the structure constants
involving two Ramond fields although we have not yet tried to do so.

{\it Note added}.  When this work was completed we discovered the paper
by Kitazawa et al.
\REF\japan{Y. Kitazawa, N. Ishibashi, A. Kato, K. Kobayashi, Y. Matsuo
and S. Odake:  Nucl. Phys. {\bf B306}(1988) 425.}[\japan]
where the $N=1$ superconformal structure constants were computed.
Our results agree with those of ref. [\japan], and provide a non-trivial
verification of the whole computation.  Our method is a manifestly
superconformally invariant generalization of the work of
Dotsenko and Fateev [\dotfatone, \dotfattwo, \dotfatthree].  The
results are complicated enough that an independent computation
of these structure constants
is not unreasonable.
We also feel that some of the technical
problems we tackled in dealing with supercontour integrals
are interesting in their own right, and they can provide
a basis for extension to other cases like for example
$N=2$ models.

\endpage

\taghead{2.}
\chapter{GENERAL PROPERTIES OF SUPERCONFORMAL THEORIES}

\def\twoone{D = {\partial \over \partial \theta} + \theta {\partial \over
\partial z} }

\def\twotwo{Z \rightarrow \wtilde Z (Z) =
 ( \wtilde z (Z), \wtilde \theta (Z) )}

\def\twothree{D = D \wtilde \theta \; \wtilde D}

\def\twofour{D \wtilde z = \wtilde \theta D \wtilde \theta}

\def\twofive{d \wtilde Z = D \wtilde \theta \; dZ}

\def\twosix{\phi(Z) dZ^{2h}}

\def\twoseven{\oint d \theta \; \theta = 1 \qquad \oint d \theta \; 1 = 0}

\def\twoeight{\oint_C dZ \; \omega (Z) = \oint_C dz
 \oint d\theta \omega(Z) =
 \oint_C dz \omega_1(z)}

\def\twonine{f(Z_1,Z_2) = \int_{Z_2}^{Z_1} dZ \; \omega(Z)}

\def\twoten{f(Z_2,Z_2) = 0 \qquad D_1 f(Z_1,Z_2) = \omega(Z_1)}

\def\twoeleven{\eqalign{ \theta_{12} & \equiv \theta_1 - \theta_2 =
\int_{Z_2}^{Z_1} dZ \cr
 Z_{12} &\equiv z_1 - z_2 - \theta_1 \theta_2 = \int_{Z_2}^{Z_1} dZ \;
 \int_{Z_2}^Z dZ' }}

\def\twotwelve{\eqalign{ f(Z_1) &= \sum_{n=0}^\infty {1 \over n!} \Zot^n
\partial_2^n
  (1+ \Tot D_2) f(Z_2) \cr
 &= f(Z_2) + \Tot D_2 f(Z_2) + z_{12} \partial_2 f(Z_2) + \dots }}

\def\twothirteen{\eqalign{ {1 \over 2 \pi i} &
 \oint_{C_2} dZ_1 \; \Zot^{-n-1} =
0 \cr
 {1 \over 2 \pi i} & \oint_{C_2} dZ_1 \; \Tot \Zot^{-n-1} = \delta_{n,0} }}

\def\twofourteen{\eqalign{ {1 \over 2 \pi i} &
 \oint_{C_2} dZ_1 \; f(Z_1) \Tot
\Zot^{-n-1} =
  {1 \over n!} \partial_2^n f(Z_2) \cr
 {1 \over 2 \pi i} & \oint_{C_2} dZ_1 \; f(Z_1) \Zot^{-n-1} =
  {1 \over n!} \partial_2^n D_2 f(Z_2) }}

\def\twofifteen{\wtilde Z = g(Z) \qquad
\wtilde z(Z) = {a z + b + \alpha \theta \over c z +
 d + \beta \theta} \qquad
\wtilde \theta(Z) = {\bar \alpha z + \bar \beta + \bar A \theta \over
 c z + d + \beta \theta}}

\def\twosixteen{\bar \alpha = {a \beta - c \alpha
 \over \sqrt{a d - b c} } \qquad
\bar \beta = {b \beta - d \alpha \over \sqrt{ad-bc} } \qquad
\bar A = \sqrt{ad-bc-3 \alpha \beta}}

\def\twoseventeen{{\rm sdet} \; g \equiv ad-bc- \alpha \beta}

\def\twoeighteen{\eqalign{ D \wtilde \theta &=
 {\sqrt{{\rm sdet} \; g} \over
cz+d+\beta \theta}
  \cr
 \wtilde Z_{12} &= {\rm sdet} \; g {\Zot \over (cz_1+d+\beta \theta_1)
  (cz_2+d+\beta \theta_2)} }}

\def\twonineteen{{\Zot Z_{34} \over \Zoth Z_{24}}}

\def\twotwenty{\eta = (\Zot \Zoth \Ztth)^{1/2}
 ( \theta_1 \Ztth + \theta_2 Z_{31}
+
 \theta_3 \Zot + \theta_1 \theta_2 \theta_3 )}

\def\twotwentyone{\eqalign{ \wtilde z(Z) &=
 {Z_{\cdot 1} \Ztth \over Z_{\cdot 3}
Z_{21} } \cr
 \wtilde \theta(Z) &= - {1 \over Z_{\cdot 3}}
 \sqrt{{\Ztth \over \Zot Z_{31}}}
  ( \theta_1 Z_{\cdot 3} + \theta_3 Z_{1 \cdot} - \theta \Zoth + \theta_1
  \theta \theta_3 ) }}

\def\twotwentytwo{\eqalign{ \hat z(\wtilde Z) &=
 \wtilde z (1+\wtilde \theta_2
\wtilde \theta) \cr
 \hat \theta(\wtilde Z) &= \wtilde \theta - \wtilde \theta_2 \wtilde z_2 }}

\def\twotwentythree{\eqalign{ \langle \prod_{i=1}^n \Phi_i(Z_i) \rangle =
  &\prod_{i<j} Z_{ij}^{-\gamma_{ij}} F(z_a, \eta_\alpha) \cr
 a=1, \dots, n-3 \qquad &\qquad\alpha=1,\dots,n-2 \cr
 \sum_{j \not= i} \gamma_{ji} = 2h_i \qquad & \gamma_{ij} = \gamma_{ji}
 \quad \gamma_{ii} = 0 \cr}}

\def\twotwentyfour{\langle \Phi_1(Z_1) \Phi_2(Z_2) \Phi_3(Z_3) \rangle =
\Zot^{-\gamma_{12}}
 \Zoth^{-\gamma_{13}} \Ztth^{-\gamma_{23}} (a+b \eta)}

\def\twotwentyfive{T(Z) = T_F(z) + \theta T_B(z)}

\def\twotwentysix{\delta_v \Phi(Z_2) = \oint_{C_2} dZ_1 v(Z_1) T(Z_1)
\Phi(Z_2)}

\def\twotwentyseven{T(Z_1) \Phi(Z_2) =
 {\Tot \over \Zot^2} h \Phi(Z_2) + {1/2
\over \Zot} D_2
 \Phi(Z_2) + {\Tot \over \Zot} \partial_2 \Phi(Z_2) + \dots}

\def\twotwentyeight{T(Z_1) T(Z_2) =
 {\hat c \over 4 \Zot^3} + {3 \Tot \over 2
\Zot^2} T(Z_2) + {1/2 \over \Zot} D_2 T(Z_2) + {\Tot \over \Zot} \partial_2
T(Z_2) + \dots}

\def\twotwentynine{T_B(z) = \sum_n L_n z^{-n-2}
 \qquad T_F(z) = \sum_n {1 \over
2} G_n z^{-n-3/2}}

\def\twothirty{\eqalign{ [L_n,\Phi(Z)] &= z^n
\left( z {\partial \over \partial
z} + (n+1)
   (h+ {1\over 2} \theta {\partial \over \partial \theta})
 \right) \Phi(Z) \cr
 [\epsilon G_{n+1/2}, \Phi(Z)] &= \epsilon z^n \left(
   z({\partial \over \partial \theta} - \theta {\partial \over \partial z})
   -2h(n+1) \theta \right) \Phi(Z) }}

\def\twothirtyone{\eqalign{ T(Z) &= -{1 \over 2} :D
 \phi \partial \phi:(Z) \cr
\phi(Z_1) \phi(Z_2) &\sim \log \Zot }}

\def\twothirtytwo{h(e^{i \alpha \phi}) = \half \alpha^2}

\def\twothirtythree{T(Z) = -\half : D \phi \partial \phi : (Z) + {i \over 2}
\alnot  D \partial \phi (Z)}

\def\twothirtyfour{\hat c = 1-2 \alnot^2 \qquad c =
 {3 \over 2} - 3 \alnot^2}

\def\twothirtyfive{h(e^{i \alpha \phi}) = \half \alpha ( \alpha -\alnot)}

\def\twothirtysix{Q_\pm = \oint dZ e^{i \alpha_\pm \phi}}

\def\twothirtyseven{\half \alpha_\pm ( \alpha_\pm -\alnot) = \half}

\def\twothirtyeight{\eqalign{ \alp + \alm &= \alnot \cr
 \alp \alm &= -1 \cr}}

 \def\twothirtynine{\alpha_\pm =
 {1 \over 2 \sqrt{2}} (\sqrt{1-\hat c} \pm
\sqrt{9-\hat c}) }

\def\twofourty{h_{m',m} = {\hat c -1 \over 16} + {1\over 8}
 (m'\alm + m\alp)^2
+ {1\over 32} (1-(-1)^{m'-m})}

\def\twofourtyone{\eqalign{ V_{m',m}(Z) &= e^{i \almm \phi(Z)} \cr
 \almm &= {1-m' \over 2} \alm + {1-m \over 2} \alp \cr
 m'-m &\equiv 0 \quad (\mod 2) }}

\def\twofourtytwo{\eqalign{ V_{m',m}(z) &= \sigma(z)
 e^{i \almm \phi_0(z)} \cr
 \almm &= {1-m' \over 2} \alm + {1-m \over 2} \alp \cr
 m'-m &\equiv 1 \quad (\mod 2) }}

\def\twofourtythree{\bar \alpha_{m',m} = \alnot - \almm = \alpha_{-m',-m}}

\def\twofourtyfour{p' \alm + p \alp = 0}

\def\twofourtyfive{\hat c = 1 -{2 (p'-p)^2 \over p p'} \qquad \alp =
\sqrt{p'/p}  \qquad \alm = -\sqrt{p/p'}}

\def\twofourtysix{h_{m',m} = {1 \over 8 p p'}[(mp'-m'p)^2
 - (p'-p)^2] + {1\over
32} (1-(-1)^{m'-m})}

\def\twofourtyseven{\langle \Phi_{m'_1,m_1}(Z_1) \Phi_{m'_2,m_2}(Z_2)
\Phi_{m'_3,m_3}(Z_3) \rangle }

\def\twofourtyeight{[m'_1,m_1] \times [m'_2,m_2] =
\sum_{|m'_1-m'_2|+1}^{\min\left({2p'-m'_1-m'_2-1\atop m'_1+m'_2-1}\right)}
\vphantom{\sum}'
\sum_{|m_1-m_2|+1}^{\min\left({2p-m_1-m_2-1\atop m_1+m_2-1}\right)}
\vphantom{\sum}' \quad[m',m] }

\def\twofourtynine{\lim_{R \rightarrow \infty} R^{2h_4} \langle \Phi_4
(R,R\eta) \Phi_3(1,0) \Phi_2(z,\theta) \Phi_1(0,0) \rangle }

\def\twofiftyone{R^{2h_4 + \alpha_4 (\alpha_1 +\alpha_2 +\alpha_3 +m\alp
+m'\alm)}}

\def\twofiftytwo{\eqalign{ & \oint_{C_1} t^\alpha
 (1-t)^\beta (z-t)^\gamma \cr
  & \oint_{C_2} t^\alpha (t-1)^\beta (t-z)^\gamma }}

\def\twofiftythree{\eqalign{ & \int_0^z t^\alpha
 (1-t)^\beta (z-t)^\gamma \cr
  & \int_1^\infty t^\alpha (t-1)^\beta (t-z)^\gamma }}

\def\twofiftyfour{\eqalign{ F(Z) & = F_0(z) + \theta F_1(z) \cr
  f(Z_1,Z_2) & = \int_{Z_2}^{Z_1} dZ F(Z) = \int_{z_2}^{z_1} dz F_1(z)
+ \theta_1 F_0(z_1) - \theta_2 F_0(z_2) }}

\def\twofiftyfive{\wtilde z = f(z) \qquad \wtilde
 \theta = \theta \sqrt{{\partial f
\over \partial z}}}

\def\twofiftysix{\int_{\wtilde Z_2}^{\wtilde Z_1} d\wtilde Z F(\wtilde Z) =
\int_{z_2}^{z_1} dz {\partial f \over \partial z}
 F_1(f(z)) + \wtilde\theta_1
F_0(\wtilde z_1) - \wtilde\theta_2 F_0(\wtilde z_2)}

In this section we collect some of the general properties of Superconformal
Field Theories (SCFT)
[\friedan].  Since there is abundant
literature on the  subject we mostly establish our notation
and review the basic formulae of the Coulomb gas formulation
of  $N=1$  SCFT
[\bkt].  We also present some
useful properties
of $SL(2|1)$-transformations.  Further details can be found in
the literature.

\section{SUPERCONFORMAL TRANSFORMATIONS, $SL(2|1)$}

We follow mainly
D. Friedan's lectures in [\friedan].
A superpoint in the complex superplane $C^{1|1}$ will be
denoted by  $Z=(z,\theta)$.  The superderivative $D$ is given by
$$
\twoone
\tag etwoone
$$
A function $f(Z,\overline{Z})$ is superanalytic if it satisfies
$\overline{D} f=0$.  In components $f=\phi + \theta\psi
+\overline{\theta}\lambda +\theta \overline{\theta} \chi$ and
 the analyticity
condition implies $\lambda=\chi=0$. Furthermore, $\phi,\psi$ are
holomorphic functions of $z$:  $f=\phi(z) + \theta\psi(z)$.
  A superconformal transformation is a change of coordinates
$$
\twotwo
\tag etwotwo
$$
under which $D$ transforms covariantly
$$
\twothree
\tag etwothree
$$
This implies
$$
\twofour
\tag etwofour
$$
A super-Riemann surface is built by gluing $C^{1|1}$ patches with
superconformal transformations.  In analogy with Conformal Field
 Theory (CFT)
[\bpz] we can introduce tensor fields.  If $dZ$ denotes the
superdifferential transforming according to
$$
\twofive
\tag etwofive
$$
under superconformal transformations, a superconformal tensor of rank $h$ is
a function $\phi(Z)$ such that
$$
\twosix
\tag etwosix
$$
is invariant under superconformal transformations.  As usual, we omit the
anti-holomorphic dependence whenever possible.  In components
$$
\phi(Z)=\phi_0(z)+\theta\phi_1(z)
$$
The standard conformal dimensions of $\phi_0,\phi_1$ are respectively
$h$ and $h+1/2$.  The analogues of abelian differentials are tensors of
type $1/2$.  For these we can define supercontour integrals.  Recalling
$$
\twoseven
\tag etwoseven
$$
one defines
$$
\twoeight
\tag etwoeight
$$
It is also possible to define line integrals
$$
\twonine
\tag etwonine
$$
according to
$$
\twoten
\tag eqtwoten
$$
For example
$$
\twoeleven
\tag etwoeleven
$$
A superanalytic function can be expanded in power series:
$$
\twotwelve
\tag etwotwelve
$$
The fundamental formulae of superconformal calculus are
$$
\twothirteen
\tag etwothirteen
$$
yielding the generalization of Cauchy's formula
$$
\twofourteen
\tag etwofourteen
$$
A special type of transformations is the fractional linear transformations.
Writing
$$
\twofifteen
\tag etwofifteen
$$
one easily solves \(etwofour) to obtain
$$
\twosixteen
\tag etwosixteen
$$
Define the superdeterminant of $g$ as
$$
\twoseventeen
\tag etwoseventeen
$$
One easily verifies the following properties:
$$
\twoeighteen
\tag etwoeighteen
$$
The supergroup $SL(2|1)$ has dimension $3|2$.  Therefore given any four
points $Z_i,i=1,2,3,4$ we can fix for example
$z_1=0,z_2=1,z_3=\infty ; \theta_1=\theta_2=0$.  In general
an $n$-point function of Neveu-Schwarz (NS) fields will depend on
$n-3|n-2$ parameters.  The harmonic ratio
$$
\twonineteen
\tag etwonineteen
$$
is $SL(2|1)$-invariant.  Given any
three points $Z_1,Z_2,Z_3$ we can construct an odd
 $\sltwo$ invariant quantity
$$
\twotwenty
\tag etwotwenty
$$
It is convenient to perform the transformation from
$Z_1,\dots,Z_4$ to $(0,0)$, $(1,0)$, $(\infty,\theta)$,
 $(z_4,\theta_4)$ in two
steps.  First we apply the transformation
$$
\twotwentyone
\tag etwotwentyone
$$
with $Z_{.j}=z-z_j-\theta \theta_j$.  Then we apply:
$$
\twotwentytwo
\tag etwentytwo
$$
Choosing $Z_1=(z_1,\theta_1), Z_2=(z_2,\theta_2), Z_3^{\epsilon}
=(z_3+\epsilon ,\theta_3), Z_4=(z_4,\theta_4)$, the application of
the previous two transformations has the desired result as
$\epsilon \rightarrow 0$.

Since the NS vacuum is $\sltwo$-invariant, we can write the general form
of the $n$-point function for NS-fields:
$$
\twotwentythree
\tag etwotwentythree
$$
In particular, the three-point function takes the form:
$$
\twotwentyfour
\tag etwotwentyfour
$$
with $\eta$ given in \(etwotwenty).  Both \(etwotwentythree),
 \(etwotwentyfour)
are direct consequences of \(etwofifteen), \(etwoeighteen).
  The coefficients
$a,b$ are the structure constants of the superconformal operator
algebra, and their computation is the main object of this paper.

\section{FREE SUPERFIELDS AND BACKGROUND CHARGE}

The generator of superconformal transformations is the super-energy-momentum
tensor
$$
\twotwentyfive
\tag etwotwentyfive
$$
$$
\twotwentysix
\tag etwentysix
$$
For primary superfields the operator product expansion (OPE) of
$T(Z)$ and $\phi(Z)$ is
$$
\twotwentyseven
\tag etwentyseven
$$
The OPE defining the super-Virasoro algebra is
$$
\twotwentyeight
\tag etwentyeight
$$
The mode expansions defining $L_n, G_n$ are
$$
\twotwentynine
\tag etwentynine
$$
where for $T_F, n\in Z+1/2$ in the NS sector and $n\in Z$ in the
Ramond (R) sector.  Using Cauchy's formula \(etwofourteen)
$$
\twothirty
\tag etwothirty
$$
A double check on \(etwotwentythree) can be obtained by writing
the $\sltwo$ generators $G_{\pm1/2},L_{\pm 1},L_0$
as super-differential operators according to (2.30) and then showing
that (2.23) is annihilated by them.

The standard value of the central charge of the Virasoro algebra is
$c=3\hat{c}/2$.  The simplest realization of
the superconformal algebra is provided by a free massless scalar superfield.
$$
\twothirtyone
\tag etwothirtyone
$$
One easily verifies that $\hat{c}=1$ and that the conformal dimension
of a vertex operator is given by
$$
\twothirtytwo
\tag etwothirtytwo
$$
The $n$-point correlators of vertex operators vanish unless the
charge neutrality condition
$\sum_i \alpha_i=0$ is satisfied.  The central charge $\hat {c}$
can be changed to any value by adding a background charge
[\bkt]
in analogy with the Virasoro case
[\dotfatone,\dotfattwo,\dotfatthree].  Define a new energy-momentum
tensor
$$
\twothirtythree
\tag etwothirtythree
$$
now
$$
\twothirtyfour
\tag etwothirtyfour
$$
The conformal dimension of a vertex operator also changes
$$
\twothirtyfive
\tag ethirtyfive
$$
and notice that both $\alpha$ and $\overline{\alpha}=\alpha_0-\alpha$
give fields
with the same conformal dimension.  As in the standard Coulomb gas
construction [\dotfatone] there are two screening fields of dimension
   $1/2$
leading after contour integration to the screening charges
$Q_{\pm}$:
$$
\twothirtysix
\tag etwothirtysix
$$
with $\alpha_{\pm}$ satisfying
$$
\twothirtyseven
\tag etwothirtyseven
$$
or
$$
\twothirtyeight
\tag etwothirtyeight
$$
$$
\twothirtynine
\tag etwothirtynine
$$
The charge neutrality condition for the correlation function of
$n$ vertex operators is changed to
$\sum_i \alpha_i =\alpha_0$.  In the NS sector the bosonic ($\phi_0$)
and fermionic ($\phi_1$) fields combine into a superfield
$\phi(Z)$.  They are both periodic around $z=0$.  In the
Ramond sector one has instead $G(e^{2\pi i}z)=-G(z)$;
$\phi_1$ is  antiperiodic, $\phi_0$ is periodic and they do not
combine to form a superfield.  Furthermore we have to take into
account the spin fields $\sigma^{\pm}(z)$ associated to
$\phi_1$.  The vertex operators in the R sector take the general
form $\sigma^{\pm}(z) e^{i\alpha \phi_0(z)}$ with conformal dimension
${1\over 16}+{1\over 2}\alpha(\alpha-\alpha_0)$.  This formula
is again invariant under $\alpha\mapsto\alpha_0-\alpha$.\hfil\break
The singular
representations of the super-Virasoro algebra are labelled by two
positive integers $m,m'\ge1$ with highest weights
$$
\twofourty
\tag etwofourty
$$
When $m'-m$ is even we have NS field, and for  $m'-m$ odd we have R fields.
In  the Coulomb gas picture the singular modules are generated by
the vertex operators.  In the NS sector
$$
\twofourtyone
\tag etwofourtyone
$$
and in the R sector
$$
\twofourtytwo
\tag etwofourtytwo
$$
The charge screening condition for a $n$-point correlator of vertex
operators is as usual $\sum_i\alpha_i=\alpha_0$
 independently of whether we have
NS or R fields.  It is useful to note that
$$
\twofourtythree
\tag etwofourtythree
$$
The minimal superconformal theories are those satisfying
$$
\twofourtyfour
\tag efourtyfour
$$
where $p',p$ are positive integers.  They are both supposed to be even
or odd, and their greatest common divisor is either $2$ or $1$.  The parity
condition $m'-m\equiv 0\;({\rm mod} \;2)$ follows from the fact that
(2.43),(2.44) taken together imply
$\overline{\alpha}_{m',m}=\alpha_{p'-m',p-m}$ and if $\alpha_{m',m}$
is in the NS sector, $\alpha_{p'-m',p-m}$ should also be a NS field.  In
the rational case (2.44):
$$
\twofourtyfive
\tag etwofourtyfive
$$
and we can choose $p'>p$.  The unitary series occurs when [\fqsthree]
$p'=p+2$,
$$
\twofourtysix
\tag efourtysix
$$
The primary fields of the minimal theories can be further restricted
to lie in the fundamental region
$1\le m'\le p'-1, 1\le m\le p-1, mp'-m'p\le 0$.  The charge assignments
(2.41) can also be derived by requiring the non-vanishing in the
four-point function of the lowest component of $V_{\alpha}$:
$\langle V_{\alpha}  V_{\alpha}  V_{\alpha}  V_{\overline{\alpha}}\rangle$.
The charge can only be screened by the insertion of $Q_+^{N_+}Q_-^{N_-}$
for the given values of $\alpha$.  A simple consequence of the Coulomb
gas representation is the computation of the fusion rules.  We only need
to write the three possible ways of screening the three-point function
$$
\twofourtyseven
\tag etwofourtyseven
$$
We can conjugate any of the fields in (2.47) .  For the rational case (2.45)
if we compute the fusion rules by counting screenings of both
$\langle V_{\alpha_1}V_{\alpha_2}V_{\alpha_3} \rangle$,
$\langle V_{\overline{\alpha_1}} V_{\overline{\alpha_2}}
 V_{\alpha_3}\rangle$,
and requiring compatibility we obtain in the NS sector $$ \twofourtyeight
\tag etwofourtyeight
$$
Where the prime in the sum means that $m',m$ jump in steps of two units.

\section{CORRELATORS IN THE COULOMB GAS REPRESENTATION}

We write down in this section the contour integral representation
of chiral correlators for minimal $N=1$ theories in the NS sector.  Using
(2.21,22) and the covariance properties of conformal superfields, we can
transform the four-point function
$\langle \Phi_4(Z_4) \Phi_3(Z_3) \Phi_2(Z_2) \Phi_1(Z_1)\rangle$
 into the form
$$
\twofourtynine
\tag etwofourtynine
$$
Representing the superfields as Coulomb gas vertex operators
$e^{i\alpha_i \phi(Z_i)}$, the generic chiral four-point function
takes the form
$$
\eqalign{ \lim_{R \rightarrow \infty} R^{2h_4} &\langle
e^{i\alpha_4\phi(R,R\eta)} e^{i\alpha_3\phi(1,0)}
e^{i\alpha_2\phi(z,\theta)}
e^{i\alpha_1\phi(0)}
  \oint \prod_{a=1}^m dZ_a \prod_{a'=1}^{m'} dZ_{a'}
e^{i\alp\phi(Z_a)} e^{i\alm\phi(Z_{a'})} \rangle \cr
  = \lim_{R \rightarrow \infty} & R^{2h_4} R^{\alpha_1 \alpha_4}
(R-1)^{\alpha_3 \alpha_4} (R-z-R\eta\theta)^{\alpha_2 \alpha_4}
\oint \prod_1^m dZ_a \prod_1^{m'} dZ_{a'} \cr
  \prod_{a=1}^m & u_a^{\alpha_1 \alp} (1-u_a)^{\alpha_3 \alp}
(z-u_a-\theta \theta_a)^{\alpha_2 \alp} \prod_{a<b}^m Z_{ab}^{\alp^2} \cr
  \prod_{a'=1}^{m'} & v_{a'}^{\alpha_1 \alm} (1-v_{a'})^{\alpha_3 \alm}
(z-v_{a'}-\theta \omega_{a'})^{\alpha_2 \alm} \prod_{a'<b'}^{m'}
Z_{a'b'}^{\alm^2} \cr
  \prod_{c,d'}^{m,m'} & Z_{cd'}^{-1}
\prod_{a=1}^m (R-u_a-R\eta\theta_a)^{\alpha_4 \alp}
\prod_{a'=1}^{m'} (R-v_{a'}-R\eta\omega_{a'})^{\alpha_4 \alm} \cr
  = z^{\alpha_1 \alpha_2} & (1-z)^{\alpha_2 \alpha_3}
(1-\eta\theta)^{\alpha_2
\alpha_4} \cr
   \oint & \prod_1^m dZ_a \prod_1^{m'} dZ_{a'}
(1-\alpha_4\alp\eta \sum_1^m\theta_a- \alpha_4\alm\eta
\sum_1^{m'}\omega_{a'}) \cr
  \prod_{a=1}^m & u_a^{\alpha_1 \alp} (1-u_a)^{\alpha_3 \alp}
(z-u_a-\theta \theta_a)^{\alpha_2 \alp} \prod_{a<b}^m Z_{ab}^{\alp^2} \cr
  \prod_{a'=1}^{m'} & v_{a'}^{\alpha_1 \alm} (1-v_{a'})^{\alpha_3 \alm}
(z-v_{a'}-\theta \omega_{a'})^{\alpha_2 \alm} \prod_{a'<b'}^{m'}
Z_{a'b'}^{\alm^2} \prod_{c,d'}^{m,m'} Z_{cd'}^{-1} }
\tag etwofifty
$$
Here $Z_a\equiv (u_a,\theta_a), Z_{a'}\equiv (v_{a'},\omega_{a'})$ and we
have used $\alpha_+ \alpha_-=-1$.  Using (2.21,22) one can add to
(2.50) the appropriate prefactors giving the four-point function for
arbitrary points $Z_i$.  In the derivation of (2.50) the charge
screening condition was crucial.  Collecting all powers of $R$
we obtain a prefactor
$$
\twofiftyone
\tag etwofiftyone
$$
where $m,m'$ are the number of $+,-$ screening charges respectively.  Using
$2h_4=\alpha_4(\alpha_4-\alpha_0)$ and the screening condition
$\alpha_1 +\alpha_2 +\alpha_3+\alpha_4+m\alpha_+ +m'\alpha_-=\alpha_0$
the exponent in (2.51) vanishes and there is no $R$-dependence in the
$R\rightarrow \infty$ limit.  In the next chapter we analyze how
 to put together
chiral conformal blocks to construct physical conformal blocks.

\section{OPEN AND CLOSED CONTOURS}

We come now to a rather delicate issue in the contour definition of
supercorrelators and normalization factors.  The typical example
in the bosonic case is given by the decoupling for a level two
null-vector [\bpz].  This yields a hypergeometric differential equation
whose solutions can be represented in terms of contour integrals around the
singular points according to standard results in the theory
of ordinary differential equations
\REF\ince{\rince}[\ince].  The regular singular points of the
hypergeometric equation can be chosen at $(0,1,\infty)$ for
convenience.  The solutions $F_1(z),F_2(z)$ can be expressed as
contour integrals around $(0,z)$ and $(1,\infty)$.  When the
hypergeometric equation is applied to the computation of
conformal blocks, the normalization of its solutions is determined
by the monodromy invariance conditions.  It is therefore convenient
to write $F_1(z),F_2(z)$ as open line integrals along the cuts
joining the singular points.  In analogy with (2.50) $F_{1,2}$ are
represented up to constants by
$$
\twofiftytwo
\tag etwofiftytwo
$$
opening the contours we obtain
$$
\twofiftythree
\tag etwofiftythree
$$
The line integral representation (2.53) is simpler to use in the computation
of monodromy matrices but it has disadvantages in determining the fusion
rules and the internal channels in a given conformal block.  For
pratical purposes we will use almost exclusively the representation
(2.53).  Notice that taking the parameters $(\alpha,\beta,\gamma)$
in the range where the integral converges, the integrands vanish at the
end points of the line integrals or they have integrable divergences.

The procedure briefly outlined in the previous paragraphs can be extended
without difficulty to multi-contour integrals in the bosonic case.  For
super-contours we have to be more careful and potential ambiguities may
show up in going from closed to open contours.  The possible existence
of integration ambiguities in super-integrals is a phenomenon encountered
in Superstring Perturbation Theory
(for a clear exposition of the problem with references to the relevant
literature, see
[\ams]).  The case at hand is a milder version of
this problem.  From the definitions (2.8,9) we can compare open and closed
contour integrals.  We explicitly compute (2.9) satisfying (2.10):
$$
\twofiftyfour
\tag etwofiftyfour
$$
The limits of integration on the right-hand side of (2.54) are determined
by the even parts of the points $Z_{1,2}$.  When we open
 a closed supercontour
we should use (2.54).  It is often unavoidable to generate a nilpotent
contribution at the endpoints $z_1,z_2$.  The projected line integral
is defined by ignoring the terms linear in $\theta_{1,2}$ in (2.54).
  In other
words it is defined as a closed contour in $\theta$ but as open in $z$.
  This
prescription is inconsistent in general.  However, in our case,
 the integrals
of the type (2.50) and similar integrals to be considered in
 later sections are
such that $\theta F_0(z)$ vanishes (possibly in the
sense of analytic continuation) at the end points of the line integrals.
Effectively this  produces a projected integration prescription which is
preserved under split superconformal transformations $$
\twofiftyfive
\tag etwofiftyfive
$$
One easily checks that under (2.55)
$$
\twofiftysix
\tag etwofiftysix
$$
Hence, if $\theta F_0(z)$ vanishes at the end points of the integration
interval, the  projected prescription will be maintained by split
transformations.  Notice that, strictly speaking, only their
 difference has to
vanish in order to preserve this prescription.  Under non-split
 transformations
this prescription has to be modified.  In later sections we will
 only need to
use split transformations and the previous arguments justify the
 use of the
projected prescription.  We will remind the reader of these
 apparently obscure
considerations  when the case arises.
\endpage

\taghead{3.}
\chapter{SUPERCONFORMAL BLOCKS}

In this section we detail the structure of super-OPE, superconformal
blocks, and their integral representation.  Then we discuss the necessary
steps to compute the structure constants of the operator algebra.  For
simplicity we start with the thermal subalgebra of fields $(1,m)$
or $(m',1)$.  The presentation is tailored closely after the papers
[\bpz],[\dotfatone, \dotfattwo] and we only emphasize the intrinsic
features of the supersymmetric case.

We learned in the previous chapter that the three-point function
depends on two arbitrary constants.  Through the relation between
three-point functions and OPE we derive that the super-OPE involves
two independent sets of structure constants.  More precisely,
$\sltwo$-invariance constrains the OPE of two NS primary superfields
to be of the form
$$
\Phi_m(Z_1) \Phi_n(Z_2) = \sum_p \Zot^{-\gamma_{mnp}} A_{mn}^{p}
[\Phi_p(Z_2)]_{even} + \Zot^{-\gamma_{mnp}-1/2} B_{mn}^{p}
[\Phi_p(Z_2)]_{odd}
\tag ethreeone
$$
where $\gamma_{mnp}=h_m+h_n-h_p$, $A_{mn}^p$ and  $B_{mn}^p$ are
the structure constants of the operator algebra, and $[\Phi_p]$ denotes a
superconformal family with all its descendant fields.  The first
representatives are
$$
\eqalign{ [\Phi_p(Z_2)]_{even} & = \Phi_p(Z_2) + {h_m-h_n+h_p \over 2h_p}
\left( \Tot D_2 \Phi_p(Z_2) + \Zot \partial_2 \Phi_p(Z_2) \right) + \dots \cr
  [\Phi_p(Z_2)]_{odd\;\,} & = \Tot \Phi_p(Z_2) + {1 \over 2h_p} \Zot D_2
\Phi_p(Z_2)  + {h_m-h_n+h_p+1/2 \over 2h_p} \Zot \Tot \partial_2 \Phi_p(Z_2)
+ \dots }
\tag ethreetwo
$$
The subindex refers to the Grassmann parity of the expansion.
  These formulae
are derived by requiring the OPE to be compatible with $\sltwo$-invariance.
The full OPE is obtained by combining the holomorphic and anti-holomorphic
contributions.

\section{THE STRUCTURE OF THERMAL FOUR-POINT FUNCTIONS}

In the Coulomb gas representation the number of screening charges determines
the Grassmann parity of the correlator.  This is a consequence of the
use of vertex operators to represent primary fields and of the integration
over the fermionic variable implied in every
screening charge super-contour.  Using (2.50) we see that to an even (resp.
odd) number of screening charges corresponds an even (resp. odd) function of
the $\sltwo$-invariant variables.
  A similar argument holds for the three-point
function.  We shall use these remarks in the decomposition of the
four-point function into superconformal blocks.

For convenience we denote the thermal primary superfields in the NS sector
by $\Phi_m(Z)=\Phi_{(1,2m+1)}(Z)$ and its conjugate field by
$\Phi_{\overline{m}}(Z)$.  The out-state is as usual
$$
\langle \Phi_m(\infty)|=\lim_{R\rightarrow \infty} R^{2h_m}\langle 0|
\Phi_m(R,R\eta)
$$
The fusion rule (2.48) reads
$$
\Phi_m \times\Phi_n =\sum_{q=|m-n|}^{{\rm min}(p-m-n-2,m+n)}
\Phi_q
$$
and the three-point function is
$$
\langle \Phi_{\bar k}(\infty) | \Phi_l(1,0) \Phi_q(0,0)
 \rangle = A_{\bar k lq}
+ B_{\bar k lq} \eta
\tag ethreethree
$$
where the number of screenings is $N=l+q-k$.

The four-point function can be expanded in terms of three-point functions
when we use the OPE:
$$
\eqalign{ G & =\langle \Phi_{\bar k}(\infty) | \Phi_l(1,0) \Phi_m(z,\theta)
\Phi_n(0,0) \rangle \cr
  & = \sum_q z^{-\gamma_{mnq}} A_{mn}^q \langle \Phi_{\bar k}(\infty) |
\Phi_l(1,0) [\Phi_q(0,0)]_{even} \rangle \cr
  & \qquad + z^{-\gamma_{mnq}-1/2} B_{mn}^q \langle \Phi_{\bar k}(\infty) |
\Phi_l(1,0) [\Phi_q(0,0)]_{odd} \rangle \cr
  & =\sum_q G_q }
\tag ethreefour
$$
where the number of screening charges is $M=l+m+n-k$.  Trading the
index $q$ for $r=q-|m-n|$ we have
$$
N=l+|m-n|+r-k\equiv M+r \quad ({\rm mod}\;2)
$$
Using the previous observation on the Grassmann parity of the three- and
four-point functions, and the expansion (3.2), we see that depending on the
values of $N,M ({\rm mod}\;2)$ four types of conformal blocks $G_q$
appear.
$$\vbox{
\eqalignno{ \noalign{\hbox{$M$ even : $\quad G_q \;$ is an even function
of $z,\eta, \theta$.}}
  \hbox{$N$ even : }\; &G_q \sim A_{\bar k lq} A_{mn}^q \;z^{-\gamma} \;(1+
\eta\theta (..) + z(..) + \dots ) \qquad \cr
  \hbox{$N\;$ odd : }\; &G_q \sim B_{\bar k lq} B_{mn}^q \;z^{-\gamma-1/2}
\;(\eta\theta + z(..) + z \eta\theta (..) + \dots ) \qquad
& (ethreefive) \cr
\noalign{\hbox{$M$ odd : $ \;\quad G_q \;$ is an odd function
of $z,\eta, \theta$.}}
  \hbox{$N$ even : }\; &G_q \sim A_{\bar k lq} B_{mn}^q \;z^{-\gamma-1/2}
\;(\theta + z\eta (..) + z\theta (..) + \dots )\qquad \cr
  \hbox{$N\;$ odd : }\; &G_q \sim B_{\bar k lq} A_{mn}^q \;z^{-\gamma}
\;(\eta + \theta (..) + z\eta (..) + \dots ) \qquad } }
$$
the ellipsis denotes constants.  The knowledge of this structure is essential
in order to normalize correctly the superconformal blocks.  In deriving
(3.5) we have chosen $\Phi_{\overline{k}}$ as the conjugate field.
  One readily
sees that taking the conjugate to be any other field leads to the
 same values
for $N,M ({\rm mod}2)$ which is all the results depend on.

\section{THERMAL SUPERCONFORMAL BLOCKS AND MONODROMY INVARIANTS}

According to (2.50) the integral representation of a chiral superconformal
block in the thermal subalgebra is
$$
\eqalign{ \jkm(a,b,c,\rho;Z) & = T \int_1^\infty dV_1 \dots \int_1^\infty
dV_{m-k} \int_0^Z dV_{m-k+1} \dots \int_0^Z dV_{m-1} \cr
  (1 + & a_1\eta\theta + a_2\eta \sum_1^{m-1} \theta_i) \prod_1^{m-1} v_i^a
\prod_1^{m-k} (v_i-1)^b (v_i-z-\theta_i\theta)^c \cr
  \prod_{m-k+1}^{m-1} & (1-v_i)^b (z-v_i-\theta\theta_i)^c \prod_{i<j}^{m-1}
V_{ij}^\rho }
\tag ethreesix
$$
where $V_i=(v_i,\theta_i)$ and
$$
\eqalign{ a & = \alpha_1 \alp \cr
  b & = \alpha_3 \alp \cr
  c & = \alpha_2 \alp }
\qquad \qquad
\eqalign{ a_1 & = -\alpha_2 \alpha_4 \cr
  a_2 & = -\alp \alpha_4 \cr
  \rho & = \alp^2 }
\tag ethreeseven
$$
For a detailed discussion of the ordering prescription $T$ and analytic
continuation we refer the reader to section four.  The contour integrals
in (3.6) are ordered as in fig. 3.1.
It is often more convenient to
work with the ordered integral
$$
\eqalign{ \ikm(a,b,c,\rho;Z) = & \int_1^\infty dV_1 \int_1^{V_1} dV_2 \dots
\int_1^{V_{m-k-1}} dV_{m-k} \cr
  & \int_0^Z dV_{m-k+1} \dots \int_0^{V_{m-2}} dV_{m-1}
\{ \hbox{as in \(ethreesix)} \} }
\tag ethreeeight
$$
which is simply related to $J^{(m)}_k$ by (see sec. 4.1 for details)
$$
\jkm(a,b,c,\rho;Z) = \lambda_{k-1}(\rho) \lambda_{m-k}(\rho)
 \ikm(a,b,c,\rho;Z)
\tag ethreenine
$$
A physical correlation function is a combination of holomorphic and
anti-holomorphic superconformal blocks
$$
G(Z,\bar Z) = \sum_{k,l=1}^m X_{kl} \jkm(Z) \overline{J_l^{(m)}(Z)}
\tag ethreeten
$$
with the requirement that it should
be monodromy invariant when we analytically continue the variable $Z$ (and
obviously $\overline{Z}$) along the curves shown in fig. 3.2.  The constants
$X_{kl}$ are determined using the monodromy properties of $J^{(m)}_k$ around
the points $(0,0),(1,0)$.  The integrals $J^{(m)}_k$  can be thought of as
solutions to the superdifferential equations obtained by
 decoupling null-vectors
in the superconformal modules.  The equations have regular singularities
at the points $(0,0),(1,0),(\infty,\eta\infty)$.  We can compute the
monodromy when we analytically continue $J^{(m)}_k$ along $C_0$ or
$C_1$ in fig. 3.2:
$$
C_i: \quad \jkm(Z) \rightarrow (g_i)_{kl} J_l^{(m)}(Z)
\tag ethreeeleven
$$
where $g_i$ are $(m-1)\times(m-1)$ matrices.  With the choice of contours
made, the matrix $g_0$ is diagonal and unitary.  Monodromy invariance
under $g_0$ requires
$$
G(Z,\bar Z) = \sum_{k=1}^m X_k \jkm(Z) \overline{\jkm(Z)}
\tag ethreetwelwe
$$
The constants $X_k$ are fixed if we require
invariance under $g_1$ as well.  For this purpose it is convenient to
use another basis  ${\wtilde J}^{(m)}_k(Z)$ with diagonal monodromy
at $(1,0)$ and expand $J^{(m)}_k$ in this new basis.  Explicitly
${\wtilde J}^{(m)}_k(Z)$ are given by the integrals
$$
\eqalign{ \tjkm(a,b,c,\rho;Z) & = T \int_0^{-\infty} dV_1 \dots
\int_0^{-\infty}
dV_{m-k} \int_1^Z dV_{m-k+1} \dots \int_1^Z dV_{m-1} \cr
  (1 + & a_1\eta\theta + a_2\eta \sum_1^{m-1} \theta_i) \prod_1^{m-1}
(1-v_i)^b \prod_1^{m-k} (-v_i)^a (z-v_i-\theta\theta_i)^c \cr
  \prod_{m-k+1}^{m-1} & v_i^a (v_i-z-\theta_i\theta)^c \prod_{i>j}^{m-1}
V_{ij}^\rho }
\tag ethreethirteen
$$
with the contours ordered as in fig.3.3.
Denoting by $\beta_{kl}$ the
coefficients in these linear expansions we get:
$$
\eqalign{ \jkm(a,b,c,\rho;Z) & = \beta_{kl}(a,b,c,\rho) \wtilde{J}_l^{(m)}
(a,b,c,\rho;Z) \cr
  \tjkm(a,b,c,\rho;Z) & = \wtilde{\beta}_{kl}(a,b,c,\rho) J_l^{(m)}
(a,b,c,\rho;Z) }
\tag ethreefourteen
$$
Rewriting $G(Z,\bar Z)$ in terms of ${\wtilde J}^{(m)}_k(Z)$ and requiring
invariance under $g_1$ yields the conditions
$$
\sum_{k=1}^m X_k \beta_{kl} \beta_{kn} = 0 \qquad l \not= n
\tag ethreefifteen
$$
As in [\dotfattwo] the solution to these constraints can be chosen as
$$
X_k = {\beta_{mm} \wtilde{\beta}_{mk} \over \wtilde{\beta}_{mm} \beta_{km}}
X_m
\tag ethreesixteen
$$
This expression is invariant under simultaneous rescaling
 of the coefficients
$X_k$.  This amounts to changing the normalization of the
 four-point function.
This freedom will be used to simplify the formulae for the
 structure constants
of the operator algebra.  Before doing this however we need to compute the
matrices $\beta_{kl}$ and to properly normalize the chiral blocks
$J^{(m)}_k$.

The computation of the matrix $\beta$ simplifies considerably once
we realize that it relies essentially on the monodromy properties of the
integrand in (3.6) and therefore we can read off $\beta$ from the
results in [\dotfattwo] with some minor modifications.  The main difference
between the ordinary chiral conformal blocks in [\dotfattwo]
and here lies in the terms $(v_i-v_j-\theta_i\theta_j)^{\rho}$
replacing the terms $(v_i-v_j)^{2\rho}$ in the integrands and the fact that
the supercontour integrals are Grassmann odd.   Thus, analytically
continuing $V_i$ over $V_j$ yields a phase $e^{\pm i\pi(\rho-1)}$
in the superconformal case, to be compared with the phase factor
 $e^{\pm i\pi2\rho}$ in the conformal case (this will be explained in
more detail in section four).  Hence to compute the matrices
$\beta$ we only need to replace $\rho$ by $(\rho -1)/2$ in the
matrices $\alpha$ of Dotsenko and Fateev [\dotfattwo].  As a non-trivial
check of our computations we evaluate independently in Appendix A
the matrix elements $\beta_{mk}$.  This will indeed be an important check
of the normalization integrals computed in section four.  The matrix
$\wtilde\beta$ can be derived from $\beta$ using the relation
$$
\tjkm(a,b,c,\rho;(z,\theta);\eta) = \epsilon_m
\jkm(b,a,c,\rho;(1-z,i\theta);
-i\eta)
\tag ethreeseventeen
$$
where $\epsilon_m=1$ ($\epsilon_m=-i$) when $m-1$ is even ($m-1$ is odd).
(3.17) is obtained by performing the change of variables
$(v_j,\theta_j)\rightarrow(1-v_j,\sqrt{-1}\theta_j)$ in (3.13).
Consequently we
find $$
\wtilde{\beta}_{kl}(a,b,c,\rho) = \beta_{kl}(b,a,c,\rho)
\tag ethreeeighteen
$$
This is not surprising if we recall the relation with the $N=0$ case
in [\dotfattwo].

We next turn to the normalization of  $J^{(m)}_k$.  These integrals possess
a singularity as $z\rightarrow 0$.  Thus we first extract the singularity
and then evaluate the resulting integral which yields an analytic function
of $z$ near $z=0$.  If in the $0\rightarrow Z$ integrals in (3.8) we
make the change of variables
$V_{m-k+i}=(v_{m-k+i},\theta_{m-k+i})\rightarrow S_i=(s_i,\omega_i)=
(v_{m-k+i}/z,\theta_{m-k+i}/\zhf)$ we obtain
$$
\eqalign{ \ikm(a,b,c,\rho;Z) & = z^{(k-1)(1/2+a+c+(k-2)\rho/2)}
\int_1^\infty \prod_1^{m-k} dV_i \int_0^1 \prod_1^{k-1} dS_i \cr
  (1+a_1\eta\theta +a_2\eta & \sum_1^{m-k}\theta_i
+a_2 z^{1/2} \eta\sum_1^{k-1}\omega_i) \prod_1^{m-k} v_i^a (v_i-1)^b
(v_i-z-\theta\theta_i)^c \prod_{i<j} V_{ij}^\rho \cr
  \prod_1^{k-1} s_i^a & (1-z s_i)^b (1-s_i-{\theta\omega_i \over z^{1/2}})^c
\prod_{i<j} S_{ij}^\rho \prod_{i,j}^{m-k,k-1}
(v_i-z s_j-\zhf\theta_i\omega_j)^\rho\ }
\tag ethreenineteen
$$
Due to the presence of $z^{1/2}$ factors, the integrand in (3.19) is still
not regular at $z=0$.  We deal with this last singularity by
 expanding all the
terms containing $\theta$ or $z^{1/2}$ in the integrand.  This yields
$$
\eqalign{ \ikm(a,b,c,\rho;Z) & = z^{\Delta_k} \int_1^\infty dV_i
\int_0^1 dS_i (1+a_1\eta\theta +a_2\eta\sum_1^{m-k}\theta_i
+a_2 z^{1/2} \eta\sum_1^{k-1}\omega_i) \cr
  (1-c\sum & {\theta\theta_i \over v_i-z}) (1-{c \over \zhf}
\sum{\theta\omega_i \over 1-s_i}) (1-\rho\zhf\sum{\theta_i\omega_j \over
v_i-z s_j}) \cr
  \prod_1^{m-k} v_i^a (v_i- & 1)^b (v_i-z)^c \prod_{i<j} V_{ij}^\rho
\prod_1^{k-1} s_i^a (1-z s_i)^b (1-s_i)^c \prod_{i<j} S_{ij}^\rho
\prod_{i,j} (v_i-z s_j)^\rho }
\tag ethreetwenty
$$
where $\Delta_k=(k-1)(a+c+(k-2){\rho \over 2} +{1\over 2})$.  Depending
on the values of $m,k$ we find by inspection the following behaviours:
$$
\vbox{
\eqalignno{ \noalign{\hbox{$(m-1)$ even : $\quad  \ikm \;$ is an
even function of $z,\eta, \theta$.}}
  \hbox{\iti) $\; (k-1)$ even : }\; &\ikm \sim
z^{\Delta_k} \; (\nkm+\eta\theta (..) + \dots ) \qquad \cr
  \hbox{\itii) $\; (k-1)\;$ odd : }\; &\ikm \sim
z^{\Delta_k-1/2} \; (\eta\theta\nkm + z(..) + \dots ) \qquad
& (ethreetwentyone) \cr
  \noalign{\hbox{$(m-1)$ odd : $\quad  \ikm \;$ is an
odd function of $z,\eta, \theta$.}}
  \hbox{\itiii) $\; (k-1)$ even : }\; &\ikm \sim
z^{\Delta_k} \; (\eta\nkm + \theta (..) + \dots ) \qquad \cr
  \hbox{\itiv) $\; (k-1)\;$ odd : }\; &\ikm \sim
z^{\Delta_k-1/2} \; (\theta\nkm + z\eta (..) + \dots ) \qquad }}
$$
where $(..)$ denote some constants.  As expected we recognize here the same
expansions as in (3.5).  The analysis of (3.5) was useful to indicate
the leading term to be normalized in the expansions of the superconformal
blocks.  Once the last $z^{1/2}$ has been taken into account by (3.21)
we can set $z=0$ in the integrand of (3.20) and the
$1\rightarrow \infty; 0\rightarrow 1$ integrals decouple.  We get for
$N^{(m)}_k(a,b,c,\rho)$:
$$
\eqalign{ \iti)\; \nkm & = \int_1^\infty \prod_1^{m-k}dV_i
v_i^{a+c+\rho(k-1)}
(v_i-1)^b \prod_{i<j} V_{ij}^\rho \int_0^1 \prod_1^{k-1}
dS_i s_i^a (1-s_i)^c
\prod_{i<j} S_{ij}^\rho \cr
  & = I(0,m-k)(a+c+\rho(k-1),b,\rho) I(k-1,0)(a,c,\rho) \cr
  & = I_{m-k}(-1-a-b-c-\rho(m-2),b,\rho) I_{k-1}(a,c,\rho) }
\tag ethreetwentytwo \iti
$$
$$
\eqalign{ \itii)\; \nkm \eta\theta & = -a_2 c \int_1^\infty
\prod_1^{m-k}dV_i
(\sum_j \eta\theta_j) v_i^{a+c+\rho(k-1)} (v_i-1)^b \prod_{i<j}
V_{ij}^\rho \cr
  & \hphantom{= -a_2c}\;
\int_0^1 \prod_1^{k-1} dS_i (\sum_j{\theta\omega_j \over 1-s_j}) s_i^a
(1-s_i)^c \prod_{i<j} S_{ij}^\rho \cr
  & = -a_2 c I^\Sigma (0,m-k)(a+c+\rho(k-1),b,\rho;\eta) \wtilde{I}^\Sigma
(k-1,0)(c,a,\rho;\theta) \cr
  & = -a_2 c \wtilde{I}^\Sigma_{m-k}(-1-a-b-c-\rho(m-2),b,\rho;\eta)
\wtilde{I}^\Sigma_{k-1}(c,a,\rho;\theta) }
\eqno(ethreetwentytwo \itii)
$$
$$
\eqalign{ \itiii)\; \nkm \eta & = a_2 \int_1^\infty \prod_1^{m-k}dV_i
(\sum_j \eta\theta_j) v_i^{a+c+\rho(k-1)} (v_i-1)^b \prod_{i<j}
V_{ij}^\rho\cr
  & \hphantom{= a_2}\; \int_0^1 \prod_1^{k-1} dS_i s_i^a
(1-s_i)^c \prod_{i<j} S_{ij}^\rho \cr
  & = a_2 I^\Sigma (0,m-k)(a+c+\rho(k-1),b,\rho;\eta) I(k-1,0)(a,c,\rho) \cr
  & = a_2 \wtilde{I}^\Sigma_{m-k}(-1-a-b-c-\rho(m-2),b,\rho;\eta)
I_{k-1}(a,c,\rho) }
\eqno(ethreetwentytwo \itiii)
$$
$$
\eqalign{ \itiv)\; \nkm \theta & = - c \int_1^\infty \prod_1^{m-k}dV_i
v_i^{a+c+\rho(k-1)} (v_i-1)^b \prod_{i<j} V_{ij}^\rho \cr
  & \hphantom{= -c}\;
\int_0^1 \prod_1^{k-1} dS_i (\sum_j{\theta\omega_j \over 1-s_j}) s_i^a
(1-s_i)^c \prod_{i<j} S_{ij}^\rho \cr
  & = -c I(0,m-k)(a+c+\rho(k-1),b,\rho) \wtilde{I}^\Sigma
(k-1,0)(c,a,\rho;\theta) \cr
  & = -c I_{m-k}(-1-a-b-c-\rho(m-2),b,\rho)
\wtilde{I}^\Sigma_{k-1}(c,a,\rho;\theta) }
\eqno(ethreetwentytwo \itiv)
$$
The symbols $I,\wtilde I, \wtilde I^{\Sigma},  I^{\Sigma}$ denote
 the different
ordering of integrals and they are analyzed in detail in section
four.  Let us assume for the time being that these integrals have been
evaluated.  We only need at this point the following properties
(see (4.72))
$$
\eqalign{ I_{2n}(\alpha,\beta,\rho) & = \hat{I}_{2n}(\alpha,\beta,\rho) \cr
  \alpha \wtilde{I}^\Sigma_{2n+1}(\alpha,\beta,\rho;\eta) & =
\eta \hat{I}_{2n+1}(\alpha,\beta,\rho) = \eta \hat{I}_{2n+1}
(\beta,\alpha,\rho) }
\tag ethreetwentythree
$$
These relations are essential in simplifying (3.22).  We use (3.23) in
(3.22ii,iv) to absorb the constant $c$.  For $a_2$ we make use of the charge
screening condition for a four-point function with $m-1$ screenings: $$
\sum_1^4 \alpha_i + (m-1)\alp = \alnot = \alp+ \alm
\tag ethreetwentyfour
$$
then
$$
a_2 = -\alpha_4\alp= -1-a-b-c-\rho(m-2)
\tag ethreetwentyfive
$$
Hence (3.22) reduces to a nice form:
$$
\nkm(\abcr) = (-1)^{m-1} \hat{I}_{m-k}(-1-a-b-c-\rho(m-2),b,\rho)
\;\hat{I}_{k-1}(a,c,\rho)
\tag ethreetwentysix
$$
valid for all values of $m$ and $k$.

Summarizing, we find that the ordered integral representation of the thermal
superconformal blocks can be written as
$$
\ikm(\abcr;Z) = \nkm(\abcr) \CFkm(\abcr;Z)
\tag ethreetwentyseven
$$
Where ${\cal F}^{(m)}_k(Z)$ are the normalized thermal superconformal blocks
$$
\CFkm(\abcr;Z) = z^{\Delta_k^{(m)}} f_k^{(m)}(\abcr;Z)
\tag ethreetwentyeight
$$
with the regular functions $f^{(m)}_k(Z)$ having the expansions:
$$
\eqalign{ \iti)\quad f_k^{(m)} & = 1 + \eta\theta (\cdots) + \dots \cr
  \itii)\quad f_k^{(m)} & = \eta\theta + z (\cdots) + \dots \cr
  \itiii)\quad f_k^{(m)} & = \eta + \theta (\cdots) + \dots \cr
  \itiv)\quad f_k^{(m)} & = \theta +z \eta(\cdots) + \dots }
\tag ethreetwentynine
$$
and $\Delta^{(m)}_k$ can be read off from (3.21).  Finally the correlation
function of four thermal vertex operators (2.50) can be written as
$$
G(Z,\bar Z) \sim \sum_1^m \skm(\abcr) \left| \CFkm(\abcr;Z) \right|^2
\tag ethreethirtya
$$
where the quantities
$$
\skm(\abcr) = X_k(\abcr) \left( \nkm(\abcr) \right)^2
\tag ethreethirtyonea
$$
contain all the information about the structure constants of the operator
algebra.  These will be evaluated in section six after the normalization
integrals are computed.

\section{GENERAL SUPERCONFORMAL BLOCKS AND MONODROMY INVARIANTS}

We now extend the analysis of four-point functions to the full algebra
of $(m',m)$ fields.  Both $+,-$ screening charges will be involved although
the same program can be carried out to the end without major differences.
  The
structure of the four-point function is identical to the thermal case (3.5)
where $M$ (resp. $N$) denotes the total number of $+$ and $-$
 screenings in the
four-point (resp. three-point) function.  The integral representation
for a chiral superconformal block is
$$
\eqalign{ \jlknm(\abcr;Z) & = T \int_1^\infty \prod_1^{n-l} dU_i \int_0^Z
\prod_{n-l+1}^{n-1} dU_i \int_1^\infty \prod_1^{m-k} dV_i \int_0^Z
\prod_{m-k+1}^{m-1} dV_i \cr
  & (1 +a_1 \eta\theta +a_2 \eta\sum_1^{n-1} \theta_i +a_3 \eta\sum_1^{m-1}
\omega_i) \prod_{i,j}^{n-1,m-1} (u_i-v_j-\theta_i\omega_j)^{-1} \cr
  \prod_1^{n-1} u_i^{a'} & \prod_1^{n-l} (u_i-1)^{b'}
(u_i-z-\theta_i\theta)^{c'}
\prod_{n-l+1}^{n-1} (1-u_i)^{b'} (z-u_i-\theta\theta_i)^{c'}
\prod_{i<j}^{n-1} U_{ij}^\rho \cr
  \prod_1^{m-1} v_i^a & \prod_1^{m-k} (v_i-1)^b (v_i-z-\omega_i\theta)^c
\prod_{m-k+1}^{m-1} (1-v_i)^b (z-v_i-\theta\omega_i)^c \prod_{i<j}^{m-1}
V_{ij}^\rho }
\tag ethreethirty
$$
where $U_i=(u_i,\theta_i),V_i=(v_i,\omega_i)$ and
$$
\eqalign{ a & = \alpha_1 \alp \cr
  b & = \alpha_3 \alp \cr
  c & = \alpha_2 \alp \cr
  \rho & = \alp^2 }
\qquad
\eqalign{ a' & = \alpha_1 \alm = -\rho' a \cr
  b' & = \alpha_3 \alm = -\rho' b \cr
  c' & = \alpha_2 \alm = -\rho' c \cr
  \rho' & = \alm^2 = \rho^{-1} }
\qquad
\eqalign{ a_1 & = -\alpha_4 \alpha_2 \cr
  a_2 & = -\alpha_4 \alm = -\rho' a_3 \cr
  a_3 & = -\alpha_4 \alp }
\tag ethreethirtyone
$$
The ordering prescription $T$ is explained in sections 4.1, 5.1 and the
contours in \(ethreethirty) are ordered according to fig. 3.4.  One can
also define the superconformal blocks $\wtilde J^{(nm)}_{lk}(a,b,c,\rho;Z)$
by a straightforward extension of (3.13).  The relation with the ordered
integrals
$$
\eqalign{ \ilknm(\abcr;Z) & = \int_1^\infty dU_1 \int_1^{U_1} dU_2 \dots
\int_1^{U_{n-l-1}} dU_{n-l} \int_0^Z dU_{n-l+1} \dots \int_0^{U_{n-2}}
dU_{n-1} \cr
  \int_1^\infty dV_1 & \dots
\int_1^{V_{m-k-1}} dV_{m-k} \int_0^Z dV_{m-k+1} \dots \int_0^{V_{m-2}}
dV_{m-1} \{ \hbox{as in \(ethreethirty)} \} }
\tag ethreethirtytwo
$$
is
$$
\jlknm(\abcr;Z) = \lambda_{l-1}(\rho') \lambda_{n-l}(\rho')
\lambda_{k-1}(\rho) \lambda_{m-k}(\rho) \ilknm(\abcr;Z)
\tag ethreethirtythree
$$
Notice that the form of the coupling terms $(u_i-v_j-\theta_i\omega_j)^{-1}$
and the relations \(ethreethirtyone) imply that we can permute the $C$ and
$S$ contours without affecting the value of the integral.  This also implies
that the monodromy properties of  $J^{(nm)}_{lk}(a,b,c,\rho;Z)$ can be
derived from those of the thermal case, and indeed, we have for the
$\beta$ matrices:
$$
\beta_{(lk)(rs)}(a,b,c,\rho) = \beta_{lr}(a',b',c',\rho') \beta_{ks}(\abcr)
\tag ethreethirtyfour
$$
This in turn implies that the coefficients $X_{ks}$ in the decomposition
of the four-point function:
$$
G(Z,\bar Z) = \sum_{k,s=1}^{n,m} X_{ks} I_{ks}(Z) \overline{I_{ks}(Z)}
\tag ethreethirtyfive
$$
also split in product of the thermal results (3.16)
$$
X_{ks} = X_k(a',b',c',\rho') X_s(\abcr)
\tag ethreethirtysix
$$
As in the conformal case, the full algebra results are not just products
of thermal quantities because of the normalization factors appearing
in the superconformal blocks \(ethreethirtytwo).  The procedure for
computing the factors $N^{(nm)}_{lk}(a,b,c,\rho)$ is identical to
the thermal case although a bit more cumbersome.  We only state some
of the necessary steps.  After a simple change of variables for the
$0\rightarrow Z$ integrals and expanding the terms containing either
$\theta$ or $z^{1/2}$ we end up with a cumbersome expression to evaluate:
$$
\eqalign{ \ilknm(Z) & = z^{\Delta_{lk}} \int_1^\infty \prod_1^{n-l} dU_i
\int_0^1 \prod_1^{l-1} dT_i \int_1^\infty \prod_1^{m-k} dV_i \int_0^1
\prod_1^{k-1} dS_i \cr
  & (1 +a_1 \eta\theta +a_2 \eta\sum_1^{n-l} \theta_i +a_2 \zhf\eta
\sum_1^{l-1} \zeta_i +a_3 \eta\sum_1^{m-k} \omega_i +a_3 \zhf\eta
\sum_1^{k-1} \nu_i) \cr
  (1-c'\sum & {\theta_i\theta \over u_i-z}) (1-{c' \over \zhf}
\sum{\theta\zeta_i \over 1-t_i}) (1-\rho'\zhf \sum{\theta_i\zeta_j
\over u_i-z t_j}) (1+\zhf\sum{\theta_i\nu_j \over u_i-z s_j}) \cr
   (1-c \sum & {\omega_i\theta \over v_i-z}) (1-{c \over \zhf} \sum
{\theta\nu_i \over 1-s_i}) (1-\rho\zhf\sum{\omega_i\nu_j \over v_i-z s_j})
(1+\zhf\sum{\zeta_i\omega_j \over z t_i-v_j}) \cr
  \prod_1^{n-l} u_i^{a'} & (u_i-1)^{b'} (u_i-z)^{c'} \prod U_{ij}^{\rho'}
\prod_1^{l-1} t_i^{a'} (1-z t_i)^{b'} (1-t_i)^{c'} \prod T_{ij}^{\rho'} \cr
  \prod_1^{m-k} v_i^a & (v_i-1)^b (v_i-z)^c \prod V_{ij}^\rho
\prod_1^{k-1} s_i^a (1-z s_i)^b (1-s_i)^c \prod S_{ij}^\rho \cr
  \prod (u_i- & z t_j)^{\rho'} \prod (v_i-z s_j)^\rho \prod (UV)_{ij}^{-1}
\prod (TS)_{ij}^{-1} \prod (u_i-z s_j)^{-1} \prod (z t_i-v_j)^{-1} }
\tag ethreethirtyseven
$$
with $T_i=(t_i,\zeta_i), S_i=(s_i,\nu_i)$ and
$$
\Delta_{lk} = (l-1)(\half+a'+c'+{\rho' \over 2}(l-2)) + (k-1)(\half+a+c+
{\rho \over 2}(k-2)) - (l-1)(k-1)
\tag ethreethirtyeight
$$
Depending on the values of $n,m,l,k$ we may have to compute integrals of the
types given in sections 5.1, 5.2.  The normalization constants are
$$
\vbox{
\eqalignno{ \noalign{\hbox{$(n+m)$ even : }}
  \;\iti) \hbox{ $(l+k)$ even : }\; &\nlknm =
I_{n-l,m-k}(-a-b-c-\rho(m-2)+n-2,b,\rho) \cr
  & \hphantom{\nlknm =}\;\; I_{l-1,k-1}(a,c,\rho) \cr
  \;\itii) \hbox{ $(l+k)$ odd : }\; &\nlknm \eta\theta =
-a_2 c' \wtilde{I}^\Sigma_{n-l,m-k}(-a-b-c-\rho(m-2)+n-2,b,\rho;\eta) \cr
   & \hphantom{\nlknm \eta\theta = -a_2 c'} \;
\wtilde{I}^\Sigma_{l-1,k-1}(c,a,\rho;\theta) & (ethreethirtynine) \cr
  \noalign{\hbox{$(n+m)$ odd : }}
  \;\itiii) \hbox{ $(l+k)$ even : }\; &\nlknm \eta =
-a_2 \wtilde{I}^\Sigma_{n-l,m-k}(-a-b-c-\rho(m-2)+n-2,b,\rho;\eta) \cr
  & \hphantom{\nlknm \eta = -a_2} \; I_{l-1,k-1}(a,c,\rho) \cr
  \;\itiv) \hbox{ $(l+k)$ odd : }\; &\nlknm \theta =
c' I_{n-l,m-k}(-a-b-c-\rho(m-2)+n-2,b,\rho) \cr
  & \hphantom{\nlknm \theta = c'} \;
\wtilde{I}^\Sigma_{l-1,k-1}(c,a,\rho;\theta) } }
$$
Using \(ethreethirtyone) the charge screening condition for a four-point
function with $(n-1)$ $-$ screenings and $(m-1)$ $+$ screenings and the
definition (5.39) of $\hat I(\alpha,\beta,\rho)$ we finally get a unified
expression for the normalization constants
$$
\nlknm(\abcr) = \hat{I}_{n-l,m-k}(-a-b-c-\rho(m-2)+n-2,b,\rho) \;
\hat{I}_{l-1,k-1}(a,c,\rho)
\tag ethreeforty
$$
Thus we can define the normalized superconformal blocks ${\cal
F}^{(nm)}_{lk}(Z)$ as
$$
\ilknm(\abcr;Z) = \nlknm(\abcr) \CFlknm(\abcr;Z)
\tag ethreefortyone
$$
where
$$
\CFlknm(\abcr;Z) = z^{\Delta_{lk}^{(nm)}} f_{lk}^{(nm)}(\abcr;Z)
\tag ethreefortytwo
$$
$f^{(nm)}_{lk}$ has an expansion similar to \(ethreetwentynine) and
$\Delta^{(nm)}_{lk}$ is defined by \(ethreethirtyeight) with a $-1/2$
correction depending on the values of $l,k,n,m$.

Finally the general correlation function of four vertex
 operators can be written
as
$$
G(Z,\bar Z) \sim \sum_{k,l} \slknm(\abcr) \left| \CFlknm(\abcr;Z) \right|^2
\tag ethreefortythree
$$
where
$$
\slknm(\abcr) = X_l(a',b',c',\rho') X_k(\abcr)
 \left( \nlknm(\abcr) \right)^2
\tag ethreefortyfour
$$
encompasses the full information on the structure constants of the operator
algebra.
\endpage

\taghead{4.}
\chapter{NORMALIZATION INTEGRALS IN THE THERMAL SERIES}

\section{CONTOUR ORDERING}

{}From the analysis of conformal blocks in the previous section we are left
with the problem of evaluating a set of normalization integrals before
we can explicitly write down the structure constants of the operator
algebra.  Since the evaluation of these integrals is involved we present
the method first in detail for the correlators in the thermal series.  The
extension of the method to the general case presents some subtleties
which require the explicit result for the thermal series. This extension
will be the subject of the next section.  As previously mentioned, the
type of analysis we pursue here is tailored after the treatment of
the conformal case in [\dotfatone,\dotfattwo,\dotfatthree].  There are
specific complications to the superconformal case as the reader
familiar with these references will see in this and the next section.

Two types of integrals can be distinguished depending on the number
of super-contours.  We may have an even or an odd number of contours.
In the even case the integrals involved are of the form
$$
\eqalign{ J(p,q) = T_0 \int_1^\infty \prod_1^q dT_i & t_i^\alpha
(t_i-1)^\beta \prod_{i>j} (t_i-t_j-\theta_i \theta_j)^\rho \cr
  \int_0^1 \prod_1^p dU_a & u_a^\alpha
(1-u_a)^\beta \prod_{a<b} (u_a-u_b-\omega_a \omega_b)^\rho \cr
  \prod_{i,a} & (t_i-u_a-\theta_i\omega_a)^\rho \qquad\qquad
\hbox{$(p+q)$ even} }
\tag efourone
$$
where $T_i=(t_i,\theta_i),U_a=(u_a,\omega_a)$ and $T_0$ is
the contour ordering prescription. The contours in (4.1) are chosen with
the ordering given in fig.4.1.  For odd integrals there are two subcases
to consider
$$
\eqalign{ J^k(n-p,p) = T_0 & \int_1^\infty \prod_{n-p+1}^n dT_i
t_i^\alpha (t_i-1)^\beta \cr
  & \int_0^1 \prod_1^{n-p} dT_i t_i^\alpha (1-t_i)^\beta \;\eta \theta_k
\;\prod_{i>j} (t_i-t_j-\theta_i \theta_j)^\rho }
\tag efourtwo
$$
and
$$
\eqalign{ \wtilde{J}^k(n-p,p) = T_0 & \int_1^\infty \prod_{n-p+1}^n
dT_i t_i^\alpha (t_i-1)^\beta \cr
  & \int_0^1 \prod_1^{n-p} dT_i t_i^\alpha 1-(t_i)^\beta \;\eta {\theta_k
\over t_k}\; \prod_{i>j} (t_i-t_j-\theta_i \theta_j)^\rho }
\tag efourthree
$$
The normalization integrals are only
$$
\eqalign{ J^\Sigma(n-p,p) & = \sum_{k=1}^n J^k(n-p,p) \cr
  \wtilde{J}^\Sigma(n-p,p) & = \sum_{k=1}^n \wtilde{J}^k(n-p,p) }
\tag efourfour
$$
The contours in (4.2,3) are ordered as in fig. 4.2.  The evaluation of
(4.2,3) is harder than (4.4).  Fortunately only the latter is
needed in our computations.  The ordering prescription is best illustrated
with fig. 4.3 and the integral
$$
J_m = T_0 \int_0^X \prod_1^m dT_i \prod_{i<j}
(t_i-t_j-\theta_i \theta_j)^\rho
\tag efourfive
$$
$T_0$ orders the $d\theta_i$'s as well as the integrand
$$
\eqalign{ T_0 \Big\lbrace \prod_1^m d\theta_i \prod_{i<j}^m
(t_i-t_j-\theta_i \theta_j)^\rho \Big\rbrace & = d\theta_1 d\theta_2 \dots
d\theta_m \prod_{i<j}^m (t_i-t_j-\theta_i \theta_j)^\rho \cr
  & \hphantom{= d\theta_1 d\theta_2 \dots d\theta_m} \hbox{ for }
t_1 > t_2 > \dots > t_m \cr
  & = - d\theta_2 d\theta_1 d\theta_3 \dots d\theta_m
(t_2-t_1-\theta_2 \theta_1)^\rho e^{-i \pi \rho} \dots \cr
  & \hphantom{= d\theta_1 d\theta_2 \dots d\theta_m} \hbox{ for }
t_2 > t_1 > t_3 > \dots > t_m }
\tag efoursix
$$
According to this definition no residual phase is
 encountered in the first case
in (4.6) and when we braid (analytically continue) $t_i$ around $t_j$ for
$i>j$ along the curves in fig. 4.3 from the region $t_i<t_j$ to
the region $t_i>t_j$ we pick up a phase $e^{-i\pi(\rho-1)}$.
  When there are
two sets of contours $(t_i,\theta_i),i=1,\dots,m; (u_a,\omega_a),
a=1,\dots,n$
with $u_a>t_i$ the odd differentials are ordered as
$d\omega_1\dots d\omega_nd\theta_1\dots d\theta_m$ for
$u_1>u_2>\dots>u_n>t_1>\dots>t_m$.  The Grassmann variable
with the greatest real
variable comes first on the left.  With this choice we can define
an ordered integral $I_m$:
$$
I_m = \int_0^X dT_1 \int_0^{T_1} dT_2 \dots \int_0^{T_{m-1}} dT_m
\prod_{i<j} (t_i-t_j-\theta_i \theta_j)^\rho
\tag efourseven
$$
and a simple argument analogous to the manipulation of
 time orderings in field
theory relates $J_m$ to $I_m$:
$$
\eqalign{ J_m(\rho) & = \lambda_m(\rho) \epsilon_m(\rho)^{-1}
I_m(\rho) \cr
  \lambda_m(\rho) = \prod_1^m \;{\siro \over s(\rom)} \qquad
\epsilon_m(\rho) & = \prod_0^{m-1} e^{i\pi\rom k} \qquad s(x) =
\sin(\pi x) }
\tag efoureight
$$
In the general case with both types of screenings we will
 simply get a factor
$\lambda_m(\rho) \lambda_{m'}(\rho')$ $\epsilon_m(\rho)^{-1}
\epsilon_{m'}(\rho)^{-1}$.  To make the formulas as simple as possible,
in some places we use another ordering prescription $T$ which differs
from $T_0$ by the presence of a residual phase $\epsilon_m$. Integrals
with this ordering are related to $I_m$ as in \(efoureight) but without
the $\epsilon_m$'s.

\section{EVEN INTEGRALS}

We begin by deriving a set of relations satisfied by (4.1).  In the region
of $(\alpha,\beta,\rho)$ making the integral converge we pull
 the top contour
labelled by $p$ from $0\rightarrow 1$ through the upper half plane.  This
leads to a relation between $J(p,q)$ and $J(p-1,q+1)$ (the part of the
deformed contour from $1$ to $\infty$) and an integral $J^{(0)}(p-1,q)$
with one contour running from $0\rightarrow -\infty$. Similarly we can
pull the bottom contour through the lower half plane to obtain another
relation between these three integrals.  Being careful with phases
and ordering prescriptions, the top contour yields
$$
J(p,q) = e^{i\pi\alpha} J^{(0)}(p-1,q) - e^{-i\pi(\rho-1)(p-1)
-i\pi\beta} J(p-1,q+1)
\tag efourninea a
$$
while pulling the bottom contour yields
$$
J(p,q) = e^{-i\pi(\rho-1)(p-1)-i\pi\alpha} J^{(0)}(p-1,q) -
e^{i\pi\beta+i\pi(\rho-1)q} J(p-1,q+1)
\eqno(efourninea b)
$$
Eliminating $J^{(0)}(p-1,q)$ implies
$$
J(p,q) = - e^{i\pi\rom(q-p+1)} \; {s(\alpha+\beta+(\rho-1)(p-1+q/2))
\over s(\alpha+\rom(p-1))} \; J(p-1,q+1)
\tag efourten
$$
Iterating this relation we can pull all contours from the region
$0\rightarrow 1$ into $1\rightarrow\infty$ contours:
$$
\eqalign{ J(n,0) & = (-1)^n\prod_{j=0}^{n-1} e^{i\pi\rom(2j-n+1)}
\;{s(\alpha+\beta+(\rho-1)(n-1-j/2))\over s(\alpha+\rom(n-1-j))}\;J(0,n)\cr
  & = \prod_{j=0}^{n-1} {s(\alpha+\beta+(\rho-1)(n-1-j/2)) \over
s(\alpha+\rom j)} \; J(0,n) \qquad\qquad \hbox{$n$ even} }
\tag efoureleven
$$
Since $\sum_0^{n-1}(2j-(n-1))=0$, the phase in \(efoureleven) reduces
to $(-1)^n=1$ because $n$ is even.  We can now change from $J(0,n)$ to
 $J(n,0)$ by a split superconformal change of variables
$$
U_i=(u_i,\omega_i) \qquad u_i = f(t_i) \qquad \omega_i = \theta_i
\sqrt{\partial_i f} \qquad D_i \omega_i = \sqrt{\partial_i f}
\tag efourseventeena
$$
with
$$
u_i = {1\over t_i} \qquad \omega_i=
{\sqrt{-1} \over t_i} \theta_i \qquad D_i \omega_i = {\sqrt{-1} \over t_i}
\tag efourtwelve
$$
$$
\eqalign{ J(0,n)(\albr) & = T_0 \int_1^\infty \prod_1^n dT_i t_i^\alpha
(t_i-1)^\beta \prod_{i>j} (t_i-t_j-\theta_i \theta_j)^\rho \cr
  & = \int_1^\infty dT_n dT_{n-1} \dots dT_1 \quad (\dots) \cr
  & = J(n,0)(-1-\alpha-\beta-\rho(n-1),\beta,\rho) }
\tag efourthirteen
$$
leading to
$$
\eqalign{ J(n,0)(\albr) = & \prod_{j=0}^{n-1} {s(\alpha+\beta+(\rho-1)
(n-1-j/2)) \over s(\alpha+\rom(n-j-1))} \cr
  & \qquad\qquad J(n,0)(-1-\alpha-\beta-\rho(n-1),\beta,\rho) }
\tag efourfourteen
$$
Furthermore, using the change of variables in (4.1)
$$
u_i = 1-t_i \qquad \qquad \omega_i = \sqrt{-1} \theta_i
\tag efourfifteen
$$
we obtain a useful symmetry
$$
J(n,0)(\albr) = J(n,0)(\beta,\alpha,\rho)
\tag efoursixteen
$$
In deriving these formulas we have to be careful in keeping track of
the signs originating in the exchange of the $d\theta $'s necessary
to bring them to the correct ordering.  We should also recall
 the remarks at
the end of section two about open versus closed contour integrals.  As in
[\dotfattwo] \(efourten) and \(efoureleven) can be thought of as analytic
continuations of the function $J(n,0)(\alpha,\beta)$ to the complex
$\alpha$-plane.  Since the integrals we are dealing with have milder
singularities than those treated in [\dotfattwo] we need not repeat
their arguments here.  The next step consists in determining the
behaviour of $J(n,0)(\alpha,\beta)$ as $\alpha\rightarrow \infty$.  This
is achieved using a split superconformal transformation in (4.1) with
$$
u_i = e^{-t_i/\alpha} \qquad\qquad D_i\theta_i = \sqrt{-\alpha}
e^{-t_i/{2\alpha}}
\tag efoureighteena
$$
Keeping track of the phases carefully we arrive at
$$
J(n,0)(\albr) \sim \alpha^{-n/2 -n\beta- \rho n(n-1)/2}
\;(C_0 + C_1 \alpha^{-1}+ \dots ) \qquad n \hbox{ even}
\tag efournineteena
$$
With a little extra work we find the asymptotic behaviour of odd integrals
like
$$
J^k(n,0) = T_0 \int_0^1 \prod_1^n dT_i t_i^\alpha
(1-t_i)^\beta \;\theta_k\;\prod_{i>j} (t_i-t_j-\theta_i \theta_j)^\rho
\qquad n \hbox{ odd}
\tag efourtwentya
$$
and $\tilde J^k(n,0)$ (replace in \(efourtwentya)  $\theta_k\rightarrow
\theta_k/t_k$).  The answer is
$$
J^k(n,0) \sim \wtilde{J}^k(n,0) \sim \alpha^{-(n+1)/2 -n\beta-
 \rho n(n-1)/2}
\;(C'_0 + C'_1 \alpha^{-1}+ \dots ) \qquad n \hbox{ odd}
\tag efourtwentyonea
$$
Using \(efourfourteen),\(efoursixteen),\(efournineteena) we can write an
Ansatz for $J_n(\alpha,\beta,\rho) = J(n,0)(\alpha,\beta,\rho)$:
$$
J_n(\albr) = \prod_{j=0}^{n-1} {\Gamma(1+\alpha+\rom j+M_j) \;
\Gamma(1+\beta+\rom j+M_j) \over \Gamma(1+\alpha+\beta+\rom(n-1+j)+N_j)}
\;\mu_n(\albr)
\tag efourseventeen
$$
{}From the reflection condition (4.14) we obtain a relation between
the integers $M$ and $N$:
$$
N_{n-1-j} + M_j = n-1
\tag efoureighteen
$$
The function $\mu(\alpha,\beta,\rho)$ is symmetric under the exchange
of $\alpha$ and $\beta$ and it satisfies
$\mu_n(\alpha,\beta,\rho)=\mu_n(-1-\alpha-\beta-\rho (n-1),\beta,\rho)$.
  We
can obtain more constraints by matching the large $\alpha$ behaviour
\(efournineteena) :
$$
\sum_{p=0}^{n-1} M_p = {n(n-2) \over 4} \qquad \qquad n \hbox{ even}
\tag efournineteen
$$
Since $\mu(\alpha,\beta,\rho)$ is analytic in $\alpha$ and behaves as
a constant in the large $\alpha$ limit, we conclude that $\mu$ is only a
function of $\rho$.  We can obtain more information about the $M_j$'s
if we require them not to depend on $n$.  Since the case $n=2$
 can be computed
explicitly, $M_0=M_1=0$, if we subtract \(efournineteen) for $n$ and $n+2$
we obtain the relation
$$
M_{2k} + M_{2k+1} = 2k
\tag efourtwenty
$$
If $[x]$ stands for the integer part of $x$, one easily checks that
$M_j=[j/2]$ solves \(efournineteen) and \(efourtwenty).  As we shall see
later this turns out to be the correct answer.  To summarize this
subsection we have learned that for even $n$
$$
\eqalign{ J_n(\albr) = & \prod_{j=0}^{n-1} {\Gamma(1+\alpha+\rom j+M_j)
\;\Gamma(1+\beta+\rom j+M_j) \over \Gamma(1+\alpha+\beta+\rom(n-1+j)
+n-1-M_j)} \;\mu_n(\rho) \cr
  & \qquad \sum_{p=0}^{n-1} M_p = {n(n-2) \over 4} }
\tag efourtwentyone
$$
Next we turn to the odd case.

\section{ODD NUMBER OF CONTOURS}

In deriving relations out of pulling contours in the odd case it will be
convenient to write one more entry in the arguments of (4.2,3).  We write
$$
\eqalign{ J^k(0,n-p,p) & =  J^k(n-p,p) \cr
  \wtilde{J}^k(0,n-p,p) & =  \wtilde{J}^k(n-p,p) }
\tag efourtwentytwo
$$
The first entry keeps track of contours running from $0\rightarrow -\infty$.
The ordering of contours is shown in fig. 4.2.
  In this case we have to extend
the work of [\dotfattwo] to reach definite answers.  The distinguished
contour $k$ in \(efourtwentytwo)  (fig. 4.2) may be any of the $n$ contours.
In pulling the top and bottom contours we have to
 be careful with the position
of $k$.  We can distinguish five cases:
$$
\vbox{ \eqalignno{ \noalign{1) $1<k<n-p$}
\noalign{   1a) Open the top contour between $0 \rightarrow 1$}
  J^{k(-)}(0,n-p,p) & = e^{i\pi\alpha} J^{k(-)}(1,n-p-1,p) &
(efourtwentythree a) \cr
  & - e^{-i\pi(\rho-1)(n-p-1)-i\pi\beta} J^{k(-)}(0,n-p-1,p+1) } }
$$
$$
\vbox{ \eqalignno{
\noalign{   1b) Open the bottom contour between $0 \rightarrow 1$}
  J^{k(-)}(0,n-p,p) & = e^{-i\pi\alpha-i\pi(\rho-1)(n-p-1)} J^{k-1(-)}
(1,n-p-1,p) & (efourtwentythree b)\cr
  & - e^{i\pi\beta+i\pi(\rho-1)p} J^{k-1(-)}(0,n-p-1,p+1) } }
$$
These formulae look cumbersome, but the meaning is simple.  By
$J^{k(-)}$ we mean that the distinguished variable $\theta_k$ corresponds
to the odd integration variable of the $k$-th contour from
$0\rightarrow 1$ in fig. 4.2 counting from the bottom. This is why
on the right-hand side of \(efourtwentythree b) the label of $J$ has
become $k-1(-)$.  By pulling the bottom contour we have changed the
numbering.  The arguments $(1,n-p-1,p)$ mean that we have one
 contour running
from $0\rightarrow -\infty$, $n-p-1$ from $0\rightarrow 1$ and
$p$ from $1\rightarrow\infty$.  Similarly, if $\theta_k$ is related to
a contour from $1\rightarrow \infty$ we label the corresponding integral
by $J^{k(+)}$.

Notice that the phases in the contour deformations do not depend on $k$.
  The
same happens in all subsequent cases.
$$
\vbox{ \eqalignno{ \noalign{2) $n-p+1< k \leq n$}
\noalign{   2a) Open the top contour between $0 \rightarrow 1$}
  J^{k(+)}(0,n-p,p) & = e^{i\pi\alpha} J^{k-1(+)}(1,n-p-1,p) &
(efourtwentyfour a)\cr
  & - e^{-i\pi(\rho-1)(n-p-1)-i\pi\beta} J^{k(+)}(0,n-p-1,p+1) } }
$$
$$
\vbox{ \eqalignno{
\noalign{   2b) Open the bottom contour between $0 \rightarrow 1$}
  J^{k(+)}(0,n-p,p) & = e^{-i\pi\alpha-i\pi(\rho-1)(n-p-1)} J^{k-1(+)}
(1,n-p-1,p) & (efourtwentyfour b)\cr
  & - e^{i\pi\beta+i\pi(\rho-1)p} J^{k-1(+)}(0,n-p-1,p+1) } }
$$
$$
\vbox{ \eqalignno{ \noalign{3) $k=1$}
\noalign{   3a) Open the top contour}
  J^{1(-)}(0,n-p,p) & = e^{i\pi\alpha} J^{1(-)}(1,n-p-1,p) &
(efourtwentyfive a) \cr
  & - e^{-i\pi(\rho-1)(n-p-1)-i\pi\beta} J^{1(-)}(0,n-p-1,p+1) } }
$$
$$
\vbox{ \eqalignno{
\noalign{   3b) Open the bottom contour}
  J^{k(-)}(0,n-p,p) & = e^{-i\pi\alpha-i\pi(\rho-1)(n-p-1)} J^{(-1)}
(1,n-p-1,p) & (efourtwentyfive b) \cr
  & - e^{i\pi\beta+i\pi(\rho-1)p} J^{n(+)}(0,n-p-1,p+1) } }
$$
$$
\vbox{ \eqalignno{ \noalign{4) $k=n-p$}
\noalign{   4a) Open the top contour}
  J^{n-p(-)}(0,n-p,p) & = e^{i\pi\alpha} J^{(-1)}(1,n-p-1,p) &
(efourtwentysix a) \cr
  & - e^{-i\pi(\rho-1)(n-p-1)-i\pi\beta} J^{n-p(+)}(0,n-p-1,p+1) } }
$$
$$
\vbox{ \eqalignno{
\noalign{   4b) Open the bottom contour}
  J^{n-p(-)}(0,n-p,p) & = e^{-i\pi\alpha-i\pi(\rho-1)(n-p-1)} J^{n-p-1(-)}
(1,n-p-1,p) & (efourtwentysix b) \cr
  & - e^{i\pi\beta+i\pi(\rho-1)p} J^{n-p-1(-)}(0,n-p-1,p+1) } }
$$
$$
\vbox{ \eqalignno{ \noalign{5) $k=n-p+1$}
\noalign{   5a) Open the top contour}
  J^{n-p+1(+)}(0,n-p,p) & = e^{i\pi\alpha} J^{n-p(+)}(1,n-p-1,p) &
(efourtwentyseven a) \cr
  & - e^{-i\pi(\rho-1)(n-p-1)-i\pi\beta} J^{n-p-1(+)}(0,n-p-1,p+1) } }
$$
$$
\vbox{ \eqalignno{
\noalign{   5b) Open the bottom contour}
  J^{n-p-1(+)}(0,n-p,p) & = e^{-i\pi\alpha-i\pi(\rho-1)(n-p-1)} J^{n-p(+)}
(1,n-p-1,p) & (efourtwentyseven b) \cr
  & - e^{i\pi\beta+i\pi(\rho-1)p} J^{n-p(+)}(0,n-p-1,p+1) } }
$$
Multiplying all the bottom contours relations by $e^{i\pi\alpha + i\pi
(\rho-1)(n-p-1)}$ and all the top contour by $e^{-i\pi\alpha}$ and
then adding the top and subtracting the bottom contours we arrive at
$$
\eqalign{ J^\Sigma(0,n-p,p) = - & e^{-i\pi\rom(n-1-2p)} \; {s(\alpha+\beta+
(\rho-1)(n-1-p/2)) \over s(\alpha+\rom(n-1-p))} \cr
  &\qquad\qquad J^\Sigma(0,n-p-1,p+1) }
\tag efourtwentyeight
$$
I  terating \(efourtwentyeight) we obtain
$$
J^\Sigma(0,n,0) = (-1) \prod_{j=0}^{n-1} {s(\alpha+\beta+
(\rho-1)(n-1-j/2)) \over s(\alpha+\rom j)} \; J^\Sigma(0,0,n)
\tag efourtwentynine
$$
The same argument can be carried out with $\wtilde J^{\Sigma}$.  The only
differences are sign changes in the $\wtilde J^{(-1)}$ term in pulling
the top contour in 4a) and in pulling the bottom contour in 3b).
  However, in
the sum the same cancellation takes place and we end up with
$$
\wtilde{J}^\Sigma(0,n,0) = (-1) \prod_{j=0}^{n-1} {s(\alpha+\beta+
(\rho-1)(n-1-j/2)) \over s(\alpha+\rom j)} \; \wtilde{J}^\Sigma(0,0,n)
\tag efourthirty
$$
If we next attempt to relate $J^{\Sigma}(0,0,n)$ with  $J^{\Sigma}(0,n,0)$
by performing the superconformal change (4.13) we find a surprise:
$$
\eqalign{ J^\Sigma(0,0,n)(\albr) & = \wtilde{J}^\Sigma(0,n,0)(-1-\alpha
-\beta-\rho(n-1),\beta,\rho) \cr
  \wtilde{J}^\Sigma(0,0,n)(\albr) & = J^\Sigma(0,n,0)(-1-\alpha
-\beta-\rho(n-1),\beta,\rho) }
\tag efourthirtyone
$$
Hence we need to do some extra work before we can write an Ansatz for the
odd integrals.

A second important difference arises in the symmetry
 with respect to the exchange
of $\alpha$ and $\beta$.  For $J^{\Sigma}$ the change of variables (4.16)
produces
$$
J^\Sigma(0,n,0)(\albr) = J^\Sigma(0,n,0)(\beta,\alpha,\rho)
\tag efourthirtytwo
$$
while for  $\wtilde J^{\Sigma}$ the correct symmetry is
$$
\alpha\wtilde{J}^\Sigma(0,n,0)(\albr) = \beta \wtilde{J}^\Sigma(0,n,0)
(\beta,\alpha,\rho)
\tag efourthirtythree
$$
To prove the last equation, write
$$
\eqalignno{ \wtilde{J}^k(\alpha,\beta) & = T_0 \int_0^1 \prod_i dU_i
u_i^\alpha (1-u_i)^\beta \;\eta {\omega_k \over u_k}\; \prod_{i<j}
U_{ij}^\rho & (efourthirtyfour) \cr
  \wtilde{J}^k(\beta,\alpha) & = T_0 \int_0^1 \prod_i dU_i
u_i^\beta (1-u_i)^\alpha \;\eta {\omega_k \over u_k} \;\prod_{i<j}
U_{ij}^\rho & (efourthirtyfive) }
$$
Implementing in the second integral the change of variables
$u_i=1-t_i, \omega_i=\sqrt{-1}\theta_i$ and keeping track of integration
variables ordering, we obtain
$$
\eqalign{ \alpha\wtilde{J}^k(\alpha,\beta)- &
\beta\wtilde{J}^k(\beta,\alpha)
=\cr
  & = T_0 \int_0^1 \prod_i dU_i u_k^\alpha (1-u_k)^\beta \eta\theta_k
({\alpha \over u_k} - {\beta \over 1-u_k}) \prod_{i \not= j} u_i^\alpha
(1-u_i)^\beta \prod_{i<j} U_{ij}^\rho \cr
  & = -T_0 \int_0^1 \prod_i dU_i u_i^\alpha (1-u_i)^\beta \; \eta\theta_k
{\partial \over \partial u_k} \prod_{i<j} U_{ij}^\rho }
\tag efourthirtysix
$$
In the second step we have integrated by parts.  The
 boundary terms vanish at
$0$ and $1$.  To see that \(efourthirtysix) vanishes note that
$Q_k=\theta_k{\partial\over \partial u_k}-{\partial\over
 \partial \theta_k}$ is
the supersymmetry generator acting on $(u_k,\theta_k)$.
  After summing over
$k$, $\sum Q_k$ annihilates $\prod_{i<j}U_{ij}^{\rho}$ because this function
is invariant under global supersymmetry.  Hence \(efourthirtysix)
 reduces to
an integral of the form
$$
- T_0 \int_0^1 \prod_i dU_i \sum_k {\partial \over \partial \theta_k}
\left( \prod_i u_i^\alpha (1-u_i)^\beta \eta \prod_{i<j} U_{ij}^\rho \right)
\tag efourthirtyseven
$$
which vanishes like all  total derivatives in the odd directions.  Finally,
the large $\alpha$ behavior was already computed in \(efourtwentyonea).

The next step in our construction is to relate
$J^{\Sigma}(n,0,0)(\alpha,\beta)$ to $J^{\Sigma}(0,0,n)(\alpha,\beta)$.
  This
is achieved by considering the contour configuration shown in fig. 4.4.
Start with $J^k(p,0,n-p)$.  Pulling the top and bottom contours on the right
and eliminating the term $J(p,1,n-p-1)$ we obtain
$$
J^\Sigma(p,0,n-p) = -e^{-i\pi\rom(n-1-2p)} \;{s(\alpha+\rom p) \over
s(\beta+\rom(n-1-p))} \;J^\Sigma(p+1,0,n-p-1)
\tag efourthirtyeight
$$
Iterating
$$
J^\Sigma(0,0,n) = (-1) \prod_{j=0}^{n-1} {s(\alpha+\rom j) \over
s(\beta+\rom j)} \;J^\Sigma(n,0,0)
\tag efourthirtynine
$$
Now by the $\sltwo$ transformation $t_i\rightarrow 1-t_i,
\theta_i\rightarrow
\sqrt{-1}\theta_i$ we can transform one integral into the other:
$$
J^k(n,0,0)(\alpha,\beta) = -J^k(0,0,n)(\beta,\alpha)
\tag efourforty
$$
which together with \(efourthirtynine) yields
$$
J^\Sigma(0,0,n)(\alpha,\beta) = \prod_{j=0}^{n-1} {s(\alpha+\rom j) \over
s(\beta+\rom j)} \;J^\Sigma(0,0,n)(\beta,\alpha)
\tag efourfortyone
$$
Using \(efourthirtyone):
$$
\eqalign{ \wtilde{J}^\Sigma(0,n,0)(-1-\alpha-\beta-\rho(n-1),\beta) =
& \prod_{j=0}^{n-1} {s(\alpha+\rom j) \over s(\beta+\rom j)} \cr
  & \; \wtilde{J}^\Sigma(0,n,0) (-1-\alpha-\beta-\rho(n-1),\alpha)}
\tag efourfortytwo
$$
Defining
$$
\gamma = -1-\alpha-\beta-\rho(n-1)
\tag efourfortythree
$$
we obtain
$$
\eqalign{ \wtilde{J}^\Sigma(0,n,0)(\gamma,\beta,\rho) = \prod_{j=0}^{n-1}
& {s(\gamma+\beta+(\rho-1)(n-1-j/2)) \over s(\beta+\rom j)} \cr
  & \quad\wtilde{J}^\Sigma(0,n,0) (\gamma,-1-\gamma-\beta-\rho(n-1),\rho)}
\tag efourfortyfour
$$
This relation can be solved as in the even case
$$
\wtilde{J}^\Sigma_n(\albr) =\eta f_n(\alpha) \prod_{j=0}^{n-1}
{\Gamma(1+\beta+\rom j+N_j) \over \Gamma(1+\alpha+\beta+\rom(n-1+j)
+n-1-N_j)} \;\mu_n(\rho)
\tag efourfortyfive
$$
Since $\wtilde J_1$ is explicitly calculable, we know that
$f_1(\alpha)=\Gamma(\alpha)$.  Imposing the symmetry relation
$\alpha \wtilde J_n(\alpha,\beta)=\beta\wtilde J_n(\beta,\alpha)$ leads to
$$
f_n(\alpha) = \prod_{j=0}^{n-1} \Gamma(1+\alpha+\rom j+\wtilde{N}_j)
\tag efourfortysix
$$
with
$$
\wtilde{N}_0 = -1, \qquad N_0 = 0, \qquad \wtilde{N}_p = N_p \;,\quad p>0
\tag efourfortyseven
$$
Summarizing:
$$
\wtilde{J}^\Sigma_n(\albr) = \eta \prod_{j=0}^{n-1}
{\Gamma(1+\alpha+\rom j+\wtilde{N}_j)\; \Gamma(1+\beta+\rom j+N_j) \over
\Gamma(1+\alpha+\beta+\rom(n-1+j)+n-1-N_j)} \;\mu_n(\rho)
\tag efourfortyeight
$$
The large $\alpha$ behaviour yields
$$
\sum_{p=0}^{n-1} N_p = {(n-1)^2 \over 4} \qquad \qquad n \hbox{ odd}
\tag efourfortynine
$$
Assuming the $N_p$'s to be independent of $n$ and using $N_0=0$ would imply
$$
N_{2k+1} + N_{2k+2} = 2k+1
\tag efourfifty
$$
Using \(efourthirtyone) we finally obtain
$$
\eqalign{ J^\Sigma_n(\albr) & = \eta \prod_{j=0}^{n-1}
{\Gamma(1+\alpha+\rom j+N_j)\; \Gamma(1+\beta+\rom j+N_j) \over
\Gamma(1+\alpha+\beta+\rom(n-1+j)+n-1-\wtilde{N}_j)} \;\mu_n(\rho) \cr
  \wtilde{J}^\Sigma_n(\albr) & = \eta \prod_{j=0}^{n-1}
{\Gamma(1+\alpha+\rom j+\wtilde{N}_j)\; \Gamma(1+\beta+\rom j+N_j) \over
\Gamma(1+\alpha+\beta+\rom(n-1+j)+n-1-N_j)} \;\mu_n(\rho) \cr
  & N_0 = 0 \qquad \wtilde{N}_0 = -1 \qquad  \sum_{p=0}^{n-1} N_p =
{(n-1)^2 \over 4} \qquad \qquad n \hbox{ odd} }
\tag efourfiftyone
$$
To complete the computation we need to determine
 the integers $M_p,N_p$ and the
functions $\mu_n(\rho)$ for both even and odd $n$.  This we do by
relating $J_{2m}$ to $J^{\Sigma}_{2m-1}$ and $J^{\Sigma}_{2m+1}$ to
 $J_{2m}$.

\section{FROM $J_{2m}$ TO  $J^{\Sigma}_{2m-1}$}

Start with $J(0,2m)$ and pull one contour into the $(0,1)$ region
as shown in fig. 4.5.  Pulling the top and bottom contours yields
$$
J(0,2m) = -e^{-i\pi\rom(2m-1)} {s(\alpha) \over s(\alpha+\beta+
\rom(2m-1))} J(1,2m-1)
\tag efourfiftytwo
$$
In the limit $\alpha=-1+\epsilon$ as $\epsilon\rightarrow 0$,
$\;s(\alpha)$ develops a zero.  Since the left-hand side does not vanish,
 the
integral on the right-hand side should develop a pole
$$
\eqalign{ J(1,2m-1) = T_0 \int_1^\infty & \prod_2^{2m}dT_i t_i^\alpha
(t_i-1)^\beta \cr
  \int_0^1 & dt_1 d\theta_1 t_1^\alpha (1-t_1)^\beta \prod_2^{2m}
(t_i-t_1-\theta_i \theta_1)^\rho \prod_{i>j=2}^{2m} T_{ij}^\rho \cr
  = T_0 \int_1^\infty & \prod_2^{2m}dT_i t_i^\alpha (t_i-1)^\beta
\int_0^1 dt_1 t_1^\alpha (1-t_1)^\beta \prod_2^{2m} (t_i-t_1)^\rho \cr
  & \qquad \left( \rho \sum_{k=2}^{2m} {\theta_k \over t_k-t_1}\right)
\prod_{i>j=2}^{2m} T_{ij}^\rho }
\tag efourfiftythree
$$
The leading divergence is obtained by expanding the integrand in powers of
$t_1/t_i$.  Multiplying by $\eta$ and taking the limit as
$\epsilon\rightarrow 0$ leads to
$$
\eta J(0,2m)(-1,\beta,\rho) = \pi\rho {e^{-i\pi\rom(2m-1)} \over
s(\beta+\rom(2m-1))} \;\wtilde{J}^\Sigma(0,2m-1)(-1+\rho,\beta,\rho)
\tag efourfiftyfour
$$
{}From the reflection formulae \(efourthirteen) and \(efourthirtyone)
we arrive at
$$
\eqalign{ \eta J(2m,0)(-\beta-\rho(2m-1),\beta,\rho) = \pi\rho &
{e^{-i\pi\rom(2m-1)} \over s(\beta+\rom(2m-1))} \cr
  & \qquad J^\Sigma(2m-1,0)
(-\beta-\rho(2m-2),\beta,\rho) }
\tag efourfiftyfive
$$
Substituting the product representations in \(efourfiftyfive), collecting
all $\beta$ dependence on one side, and dropping the $\eta$ factor:
$$
\eqalign{ & {\prod_0^{2m-1} \Gamma(1-\beta-\rho(2m-1)+\rom p+M_p)
\Gamma(1+\beta+\rom p + M_p) \over \prod_0^{2m-2} \Gamma(1-\beta-
\rho(2m-1)+\rom p+N_p) \Gamma(1+\beta+\rom p + N_p)} \cr
  &\quad\times {s(\beta+\rom(2m-1)) \over \pi} = \rho e^{-i\pi\rom(2m-1)}
{\mu_{2m-1} \over \mu_{2m}} \times \cr
  & {\prod_0^{2m-1}
\Gamma(1-\rho(2m-1)+(\rho-1)(2m-1-p/2)+2m-1-M_p) \over \prod_0^{2m-2}
\Gamma(1-\rho(2m-1)+(\rho-1)(2m-2-p/2)+2m-2-\wtilde{N}_p)} }
\tag efourfiftysix
$$
To cancel $s(\beta+\rom(2m-1))$ on the left-hand side
we take the $(2m-1)$-th term in the numerator and find that only for
$M_{2m-1}=m-1$ is there a cancellation.  Notice that the right-hand side of
\(efourfiftysix) is independent of $\beta$, and therefore all
$\beta$-dependence should disappear.  By inspection one finds that
$$
M_{2m-1} = m-1, \qquad \qquad M_p =N_p
\tag efourfiftyseven
$$
is the only way to make \(efourfiftysix) $\beta$-independent.  Once the
$\beta$-dependence is cancelled, we obtain a recursion relation
between $\mu_{2m}$ and $\mu_{2m-1}$:
$$
\mu_{2m}(\rho) = (-1)^m e^{-i\pi\rom(2m-1)} {\rho \over 2}
{\prod_0^{2m-1} \Gamma(1-\rom p-M_p) \over \prod_0^{2m-2}
 \Gamma(1-\rho -\rom
p-\wtilde{N}_p) } \mu_{2m-1}(\rho)
\tag efourfiftyeight
$$

\section{RELATING $J_{2m+1}$ TO $J_{2m}$}

As in the previous section we start with $\wtilde
J^{\Sigma}(0,2m+1)(\alpha,\beta,\rho)$ and pull one contour into the
$0\rightarrow 1$ region.  We obtain
$$
\wtilde{J}^\Sigma(0,2m+1) = -e^{-i\pi(\rho-1)m} {s(\alpha)\over
 s(\alpha+\beta+
(\rho-1)m)} \;\wtilde{J}^\Sigma(1,2m)
\tag efourfiftynine
$$
Looking at the $\wtilde J^k$ component in $\wtilde J^{\Sigma}$ we easily
learn that the leading $\alpha\rightarrow 0$ singularity
in $\wtilde J^{\Sigma}$ comes from  $\wtilde J^1$.  Taking
$\alpha=\epsilon$ and letting $\epsilon\rightarrow 0$ we arrive at
$$
\wtilde{J}^\Sigma(0,2m+1)(0,\beta,\rho) = \pi\eta
{e^{-i\pi(\rho-1)m} \over s(\beta+(\rho-1)m)}
 \; J(0,2m)(\rho,\beta,\rho)
\tag efoursixty
$$
This yields a relation between $\wtilde J^{\Sigma}(2m+1,0)$ and
$J(2m,0)$.  Following the steps of the previous section
leads to $M_{2m}=m$.  Now we can determine $M_p$ to be
$$
M_p = \left[{p \over 2} \right]
\tag efoursixtyone
$$
and the recursion relations become
$$
\eqalign{ \mu_{2m}(\rho) & = (-1)^m e^{-i\pi\rom(2m-1)} {\rho \over 2}
{\gomro \over \Gamma(1-m\rho)} \;\mu_{2m-1}(\rho) \cr
  \mu_{2m+1}(\rho) & = (-1)^{m+1} e^{-i\pi(\rho-1)m}
{\gomro \over \Gamma(\half-\rho{2m+1 \over 2})} \;\mu_{2m}(\rho) }
\tag efoursixtytwo
$$
Together, they completely determine $\mu_n(\rho)$
$$
\eqalignno{ \mu_{2m}(\rho) & = e^{-i\pi(\rho-1)m(m-1/2)}
 \left({\rho \over 2}
\right)^m {\gomro^{2m} \over \prod_1^{m} \Gamma(1-\rho p) \Gamma(\half
-\rho(p-\half))} & (efoursixtythree a) \cr
  \mu_{2m+1}(\rho) & = -e^{-i\pi(\rho+1)m(m+1/2)} \left({\rho \over 2}
\right)^m {\gomro^{2m} \over \prod_1^{m} \Gamma(1-\rho p) \Gamma(\half
-\rho(p+\half))} & (efoursixtythree b) }
$$
These two formulae can be combined into a single one
$$
\mu_n(\rho) = (-1)^n e^{i\pi M_n/2} e^{-i\pi\rho{n(n-1) \over 4}} \left(
{\rho \over 2} \right)^{M_n} {\gomro^n \over \prod_1^{n} \Gamma(1-
\rop p + M_p)}
\tag efoursixtyfour
$$
Finally we collect the formulae derived in this section for easy reference
$$
\eqalignno{
  J_n(\albr) & = \hphantom{\eta} \prod_{p=0}^{n-1}
{\Gamma(1+\alpha+\rom p+M_p)\;
\Gamma(1+\beta+\rom p+M_p) \over \Gamma(1+\alpha+\beta+\rom(n-1+p)
+n-1-M_p)} \;\mu_n(\rho) \cr
  J^\Sigma_n(\albr) & = \eta \prod_{p=0}^{n-1}{\Gamma(1+\alpha+\rom p+M_p)\;
\Gamma(1+\beta+\rom p+M_p) \over \Gamma(1+\alpha+\beta+\rom(n-1+p)
+n-1-\wtilde{M}_p)} \;\mu_n(\rho) \cr
  \wtilde{J}^\Sigma_n(\albr) & = \eta \prod_{p=0}^{n-1}
{\Gamma(1+\alpha+\rom p+\wtilde{M}_p)\;
\Gamma(1+\beta+\rom p+M_p) \over \Gamma(1+\alpha+\beta+\rom(n-1+p)
+n-1-M_p)} \;\mu_n(\rho) & (efoursixtyfive) \cr
  \mu_n(\rho) & = (-1)^n e^{i\pi M_n/2} e^{-i\pi\rom{n(n-1) \over 4}} \left(
{\rho \over 2} \right)^{M_n} {\gomro^n \over \prod_1^{n} \Gamma(1-
\rop p + M_p)} \cr
  & \qquad \wtilde{M}_0 = -1, \quad M_p = \left[{p \over 2} \right],
\quad \wtilde{M}_{p>0} = M_{p>0} }
$$
It is useful to introduce a new function $\hat J_n$ defined by
$$
\hat{J}_n(\albr) = \prod_{p=0}^{n-1} {\Gamma(1+\alpha+\rom p+M_p)\;
\Gamma(1+\beta+\rom p+M_p) \over \Gamma(1+\alpha+\beta+\rom(n-1+p)
+n-1-M_p)} \;\mu_n(\rho)
\tag efoursixtysix
$$
Then
$$
\eqalign{ J_n(\albr) & = \hat{J}_n(\albr) \qquad n \hbox{ even} \cr
  \alpha\wtilde{J}^\Sigma_n(\albr) & = \eta \hat{J}_n(\albr)
\qquad n \hbox{ odd} }
\tag efoursixtyseven
$$
If one works instead with path ordered integrals, the relation between
$I$- and $J$-integrals is given in (4.8).  We may as well introduce the
functions $\hat I_n(\alpha,\beta,\rho)$ by
$$
\eqalign{ \hat{J}_n(\albr) & = \lambda_n(\rho) \epsilon_n(\rho)^{-1}
\hat{I}_n(\albr) \cr
  & \lambda_n(\rho) = \prod_1^n \;{\siro \over s(\rom)} \qquad
\epsilon_n(\rho) = \prod_0^{n-1} e^{i\pi\rom k} }
\tag efoursixtyeight
$$

\endpage

\taghead{5.}
\chapter{NORMALIZATION INTEGRALS: THE GENERAL CASE}

We now extend the arguments of the previous section to the case when
we have both $+,-$ screening charges.  There are some important
differences in the determination of the integers $M,N$ appearing
in the arguments of the $\Gamma$-functions, but many of the arguments
can be translated directly from the thermal case.  We therefore present
less details than in the previous section.  We begin once again with
the case of even integrals.

\section{EVEN INTEGRALS}

We want to evaluate
$$
\eqalign{ \jnm(\albr) = & T_0 \int_{C_i} \prod_{i=1}^n dT_i
\int_{S_j} \prod_{j=1}^m dS_j \prod_1^n t_i^{\alpha'} (1-t_i)^{\beta'}
\prod_{i<j}^n T_{ij}^{\rho'} \cr
  & \prod_1^m s_i^\alpha (1-s_i)^\beta \prod_{i<j}^m S_{ij}^\rho
\prod_{i,j}^{n,m} (t_i-s_j-\theta_i\omega_j)^{-1} \qquad (n+m)
\hbox{ even} }
\tag efiveone
$$
with $T_i=(t_i,\theta_i),S_j=(s_j,\omega_j)$ and the integration
contours appear in fig. 5.1.  With the notation of section two,
$$
\rho' = {1\over \rho} = \alm^2 = {1 \over \alp^2} \qquad
\alpha' = -\rho' \alpha \qquad \beta' = -\rho' \beta
\tag efivetwo
$$
The ordering prescription is as in the previous chapter, and we should
notice that the coupling terms $(t_i-s_j-\theta_i\omega_j)^{-1}$
do not contribute to the monodromy if we include the signs
coming from the exchange of $dT_i$ and $dS_j$.  One can check as in
[\dotfattwo] that exchanging the $C$ and $S$ contours does not change the
answer. This implies that the monodromies of the conformal blocks
will be given as a product of the monodromy matrices for the thermal
integrals obtained by ignoring the coupling terms.  In all the contour
pulling manipulations the $C_i$ and $S_j$ contours do not feel each other.
To compute (5.1) we define first the integrals
$J\J {p'} {q'} p q(\alpha,\beta,\rho)$ as shown in fig. 5.2.

By opening the top $C_{p'}$ and the bottom $C_1$ contours we can decrease
$p'$ by one unit and increase $q'$ by one unit.  In this way we
can move all the $0\rightarrow 1$ contours $p'$ to $1\rightarrow\infty$
contours.  After we are finished with the $C$-type contours we apply the
same procedure to the $S$-type contours.  The final result is
$$
\eqalign{ J\J n 0 m 0 (\albr) = & (-1)^{n+m} \prod_0^{n-1} {s(\alpha'+
\beta'+(\rho'-1)(n-1-i/2)) \over s(\alpha'+\ropm i)} \cr
  & \prod_0^{m-1} {s(\alpha+\beta+(\rho-1)(m-1-i/2)) \over s(\alpha
+\rom i)} J\J 0 n 0 m (\albr) }
\tag efivethree
$$
The matrix label $\j {a'} {b'} a b$ means that there are
$a'$ (resp. $a$) $Q_-$ (resp. $Q_+$) contours from $0\rightarrow 1$
and
$b'$ (resp. $b$) $Q_-$ (resp. $Q_+$) contours from $1\rightarrow\infty$.
Next we use the split superconformal transformation
$t_i\rightarrow 1/t_i, s_i\rightarrow 1/s_i$, and keeping in mind the
remarks in section 2.4 we obtain
$$
\eqalign{ \jnm & (\albr) = \prod_0^{n-1} {s(\alpha'+\beta'+
(\rho'-1)(n-1-i/2)) \over s(\alpha'+\ropm i)}\cr
  & \prod_0^{m-1} {s(\alpha+\beta+(\rho-1)(m-1-i/2))
\over s(\alpha+\rom i)}
\; \jnm(-1-\alpha-\beta-\rho(m-1)+n,\beta,\rho) }
\tag efivefour
$$
This reflection property suggests the Ansatz
$$
\eqalign{ \jnm(\albr) = \prod_0^{n-1} & {\Gamma(1+\alpha'+\ropm j+M'_j)\;
\Gamma(1+\beta'+\ropm j+M'_j) \over \Gamma(1+\alpha'+\beta'+\rho'(n-1)
-m-\ropm j-M'_j)} \cr
  & \prod_0^{m-1} {\Gamma(1+\alpha+\rom j+M_j) \;\Gamma(1+\beta+\rom j+
M_j) \over \Gamma(1+\alpha+\beta+\rho(m-1)-n-\rom j-M_j)}\;\mu_{nm}(\rho) }
\tag efivefive
$$
With foresight we write $\mu_{nm}(\rho)$ without any dependence on $\alpha,
\beta$.  The Ansatz \(efivefive) is symmetrical under the exchange of
$\alpha$ and $\beta$ because the original integral had this symmetry.
Matching the large $\alpha$ behaviour of (5.1) and (5.5) leads to a
constraint on the integers $M'_p,M_p$
$$
2\sum_{p=0}^{m-1} M_p + 2\sum_{p=0}^{n-1} M'_p = {m(m-2) \over 2} +
{n(n-2) \over 2} - nm
\tag efivesix
$$
Before analyzing the case $(n+m)$ odd, we can relate the even to the odd
case as we did in the thermal case.  Starting with
$J\J 0 n 0 m$ and pulling one $S$-contour into the
$(0,1)$ region we obtain
$$
J\J 0 n 0 m = - e^{-i\pi\rom(m-1)} \;{s(\alpha) \over s(\alpha+\beta+\rom
(m-1))}\; J\J 0 n 1 {m-1}
\tag efiveseven
$$
Explicitly
$$
\eqalign{ J\J 0 n 1 {m-1} & = T_0 \int_1^\infty \prod_1^n dT_i
t_i^{\alpha'} (t_i-1)^{\beta'} \prod_{i<j} T_{ij}^{\rho'} \cr
  & \int_1^\infty \prod_2^m dS_i s_i^\alpha (s_i-1)^\beta \int_0^1 dS_1
s_1^\alpha (1-s_1)^\beta \prod_{i<j} S_{ij}^\rho
\prod_{i,j} (t_i-s_j-\theta_i\omega_j)^{-1} }
\tag efiveeight
$$
As a consequence of the $T_0$-ordering prescription, $s_1$ is
always smaller than all the other variables.  Since we are interested
in the limit $\alpha=-1+\epsilon$ as $\epsilon\rightarrow 0$, we can
expand in powers of $s_1$ and pick up the pole term in $\epsilon$
which cancels the zero from $s(\alpha)$ in (5.7).  Multiplying
by the odd $\eta$ variable we obtain after some simple manipulations
$$
\eta J\J 0 n 0 m(-1,\beta,\rho) = \pi {e^{-i\pi\rom(m-1)} \over
s(\beta+\rom(m-1))} \; \wtilde{J}^\Sigma \J 0 n 0 {m-1}(-1+\rho,
\beta,\rho)
\tag efivenine
$$
(for the definition of the odd integral $\wtilde J^{\Sigma}$ see
below).  The same argument works for the $C$-contours.  Now we set
$\alpha '=-1+\epsilon$, take the small $\epsilon$ limit and
obtain
$$
\eta J\J 0 n 0 m(\rho,\beta,\rho) = -\pi\rho' {e^{-i\pi\ropm(n-1)} \over
s(\beta'+\ropm(n-1))} \; \wtilde{J}^\Sigma \J 0 {n-1} 0 m(-1+\rho,
\beta,\rho)
\tag efiveten
$$
Using (5.3) and a similar formula for $\wtilde J^{\Sigma}$ to be derived
in the next subsection we obtain
$$
\eta \jnm(-\beta-\rho(m-1)+n,\beta,\rho) = \pi {e^{-i\pi\rom(m-1)} \over
s(\beta+\rom(m-1))} \; \wtilde{J}^\Sigma_{n,m-1}(-\beta-\rho(m-1)+n,
\beta,\rho)
\tag efiveeleven
$$
$$
\eta \jnm(-1-\beta-\rho m+n,\beta,\rho) = -\pi\rho' {e^{-i\pi\ropm(n-1)}
 \over
s(\beta'+\ropm(n-1))} \; \wtilde{J}^\Sigma_{n-1,m}(-1-\beta-\rho m+n,
\beta,\rho)
\tag efivetwelve
$$
Equations (5.11,12) will allow us to obtain recursion relations
for $\mu_{n,m}(\rho)$.

\section{ODD INTEGRALS}

We consider next the $(n+m)$ odd case.  We define two types of
integrals  $ J^{\Sigma}_{nm}$, $\wtilde J^{\Sigma}_{nm}$.
$$
\eqalign{ \jnms(\albr) = & T_0 \int_{C_i} \prod_{i=1}^n dT_i
\int_{S_j} \prod_{j=1}^m dS_j \left( \rho\eta \sum_1^m \omega_k -
\eta\sum_1^n \theta_k \right)
\prod_1^n t_i^{\alpha'} (1-t_i)^{\beta'}
\prod_{i<j}^n T_{ij}^{\rho'} \cr
  & \prod_1^m s_i^\alpha (1-s_i)^\beta \prod_{i<j}^m S_{ij}^\rho
\prod_{i,j}^{n,m} (t_i-s_j-\theta_i\omega_j)^{-1} \qquad\qquad (n+m)
\hbox{ odd} }
\tag efivethirteen
$$
The $\rho$ factor appearing in the sums can be understood from the
normalization of the conformal blocks.  The contours $C_i,S_i$ are as shown
in fig. 5.1.  Similarly we define
$$
\tjnms(\albr) = T_0 \int_{C_i} \prod_{i=1}^n dT_i
\int_{S_j} \prod_{j=1}^m dS_j \left( \rho\eta \sum_1^m {\omega_k \over
s_k} - \eta\sum_1^n {\theta_k \over t_k} \right) \; \{ \hbox{same as
in \(efivethirteen)} \}
\tag efivefourteen
$$
For contour manipulations it is convenient to define
$J^{k'}\JJ 0 {n-p} p 0 {m-q} q$ and\hfill\break
$J^{k}\JJ 0 {n-p} p 0 {m-q} q$ as in fig. 5.3.
The superindex $k'$ indicates that the factor $\theta_k$
($\theta_k/t_k$) belongs to the $C$-contours, and $k$ that
it belongs to the $S$ contours.  The matrix of labels
$\JJ {a'} {b'} {c'} a b c$ counts contours.  The first column
indicates the contours from $0\rightarrow-\infty$.  The second column
counts contours from $0\rightarrow 1$ and the last column from
$1\rightarrow\infty$.  The first row refers to $Q_-$-contours and
the second to $Q_+$-contours.  The arguments leading from
\(efourtwentythree) to \(efourthirty) can be repeated here for both
$J^{\Sigma}$ and $\tilde J^{\Sigma}$.  Using the definitions of
 fig. 5.3. we
can write $J^{\Sigma}_{nm}$ as
$$ \jnms = \rho \sum_{k=1}^m J^k\JJ 0 n 0 0 m 0
- \sum_{k'=1}^n J^{k'}\JJ 0 n 0 0 m 0 \tag efivesixteen
$$
Repeating \(efourtwentythree)$-$\(efourthirty) in the present context is
more cumbersome and leads to
$$
\eqalign{ \js\JJ 0 n 0 0 m 0(\albr) & = (-1)^{n+m} \prod_{j=0}^{n-1}
{s(\alpha'+\beta'+(\rho'-1)(n-1-j/2)) \over s(\alpha'+\ropm j)} \cr
  \prod_{j=0}^{m-1} & {s(\alpha+\beta+(\rho-1)(m-1-j/2)) \over
s(\alpha+\rom j)} \; \js\JJ 0 0 n 0 0 m(\albr) }
\tag efiveseventeen
$$
The same relation holds for $\wtilde J^{\Sigma}$.  The change
$t_i\rightarrow1/t_i; s_i\rightarrow 1/s_i$ mixes
$J^{\Sigma}$ and $\wtilde J^{\Sigma}$:
$$
\eqalign{ \js\JJ 0 0 n 0 0 m(\albr) & = \tjs\JJ 0 n 0 0 m 0(-1-\alpha
-\beta-\rho(m-1)+n,\beta,\rho) \cr
  \tjs\JJ 0 0 n 0 0 m(\albr) & = \js\JJ 0 n 0 0 m 0(-1-\alpha
-\beta-\rho(m-1)+n,\beta,\rho) }
\tag efiveeighteen
$$
As in the thermal case we can pull the $1\rightarrow\infty$
to $0\rightarrow-\infty$ contours
$$
\eqalign{ \js\JJ 0 0 n 0 0 m(\albr) & = (-1) \prod_{j=0}^{n-1}
{s(\alpha'+\ropm j) \over s(\beta'+\ropm j)} \prod_{j=0}^{m-1}
{s(\alpha+\rom j) \over s(\beta+\rom j)} \cr
  & \qquad \js\JJ n 0 0 m 0 0(\albr) }
\tag efivenineteen
$$
Now changing the variables $t_i\rightarrow 1-t_i;
 s_i\rightarrow 1-s_i$ yields
$$
\js\JJ n 0 0 m 0 0(\albr) = -\js\JJ 0 0 n 0 0 m(\beta,\alpha,\rho)
\tag efivetwenty
$$
This identity together with \(efiveeighteen,efivenineteen) implies
$$
\eqalign{ \tjnms(\gamma,-1-\alpha-\gamma-\rho(m-1)+n, & \rho) =
\prod_{j=0}^{n-1} {s(\alpha'+\ropm j) \over s(\alpha'+\gamma'+\rho'(n-1)
-m-\ropm j)} \cr
  \prod_{j=0}^{m-1} & {s(\alpha+\rom j) \over s(\alpha+\gamma
+\rho(m-1)-n-\rom j)} \;\tjnms(\gamma,\alpha,\rho) }
\tag efivetwentyone
$$
As in the thermal case we can show that
$$
\alpha\tjnms(\albr) = \beta\tjnms(\beta,\alpha,\rho)
\tag efivetwentytwo
$$
This together with \(efivetwentyone) allows us to write an Ansatz for
$\wtilde J^{\Sigma}_{nm}$ and $J^{\Sigma}_{nm}$. The case
$n+m=1$ can be computed explicitly.  Introducing the integers $N_i,N'_i$,
$\wtilde{N}_i, \wtilde{N}'_i$
we obtain the Ans\"atze
$$
\eqalign{ \tjnms(\albr) = \eta \prod_0^{n-1} & {\Gamma(1+\alpha'+\ropm j+
\wtilde{N}'_j)\;
\Gamma(1+\beta'+\ropm j+N'_j) \over \Gamma(1+\alpha'+\beta'+\rho'(n-1)
-m-\ropm j-N'_j)} \cr
  & \prod_0^{m-1} {\Gamma(1+\alpha+\rom j+\wtilde{N}_j) \;\Gamma(1+\beta+
\rom j+N_j) \over \Gamma(1+\alpha+\beta+\rho(m-1)-n-\rom j-N_j)}\;
\mu_{nm}(\rho) }
\tag efivetwentyfour
$$
and
$$
\eqalign{ \jnms(\albr) = \eta \prod_0^{n-1} & {\Gamma(1+\alpha'+\ropm j+
N'_j)\;
\Gamma(1+\beta'+\ropm j+N'_j) \over \Gamma(1+\alpha'+\beta'+\rho'(n-1)
-m-\ropm j-\wtilde{N}'_j)} \cr
  & \prod_0^{m-1} {\Gamma(1+\alpha+\rom j+N_j) \;\Gamma(1+\beta+
\rom j+N_j) \over \Gamma(1+\alpha+\beta+\rho(m-1)-n-\rom j-\wtilde{N}_j)}\;
\mu_{nm}(\rho) }
\tag efivetwentyfive
$$
Matching the large $\alpha$ behaviour we obtain a relation for the integers
$N_p,N'_p,\wtilde{N_p},\wtilde{N}'_p$:
$$
\sum_0^{n-1} \wtilde{N}'_p + N'_p \;+\; \sum_0^{m-1} \wtilde{N}_p + N_p=
-\half + {n(n-2) \over 2} + {m(m-2) \over 2} -nm
\tag efivetwentysix
$$
Finally we reduce to an even number of contours by taking $\alpha=\epsilon$
or $\alpha'=\epsilon$, $\epsilon\rightarrow 0$ as in the previous
subsection.  Omitting the details, the results are
$$
\eqalign{ \tjs\J 0 n 0 m(0,\beta,\rho) & = \eta\pi\rho {e^{-i\pi\rom(m-1)}
\over s(\beta+\rom(m-1))} \; J\J 0 n 0 {m-1}(\rho,\beta,\rho) \cr
  \tjs\J 0 n 0 m(0,\beta,\rho) & = -\eta\pi {e^{-i\pi\ropm(n-1)}
\over s(\beta'+\ropm(n-1))} \; J\J 0 {n-1} 0 m(-1,\beta,\rho) }
\tag efivetwentyseven
$$
Equivalently, using \(efiveeighteen),
$$
\eqalign{ \jnms(-1-\beta-\rho(m-1)+n,\beta,\rho) = & \eta\pi\rho
{e^{-i\pi\rom(m-1)} \over s(\beta+\rom(m-1))} \cr
&  J_{n,m-1}(-1-\beta-\rho(m-1)+n,\beta,\rho) \cr
  \jnms(-1-\beta-\rho(m-1)+n,\beta,\rho)  = &-\eta\pi
{e^{-i\pi\ropm(n-1)} \over s(\beta'+\ropm(n-1))} \cr
  &  J_{n-1,m}(-1-\beta-\rho(m-1)+n,\beta,\rho) }
\tag efivetwentyeight
$$

\section{COMPUTATION OF $\mu_{n,m}(\rho)$}

To complete the computation we have to determine the integers
$M_p,N_p$, etc.  This can be done by using the recursion relations
established in the two previous subsections.  There is, however, a simpler
method of obtaining the same answer.  Consider first the even case
$$
\eqalign{ \jnm = \prod_0^{n-1} & {\Gamma(1+\alpha'+\ropm p+M'_p)\;
\Gamma(1+\beta'+\ropm p+M'_p) \over \Gamma(1+\alpha'+\beta'+\rho'(n-1)
-m-\ropm p-M'_p)} \cr
  & \prod_0^{m-1} {\Gamma(1+\alpha+\rom p+M_p) \;\Gamma(1+\beta+\rom p+
M_p) \over \Gamma(1+\alpha+\beta+\rho(m-1)-n-\rom p-M_p)}\;\mu_{nm}(\rho) }
$$
The integers $N_p,M_p,N'_p,M'_p$ are independent of $\rho$.  Hence if we
take the limit $\rho\rightarrow -1$ (still within the domain of definition
of the integrals if $\alpha,\beta >0$), the original integral (5.1)
becomes
$$
\jnm = T_0 \int_0^1 \prod_1^{n+m} dT_i u_i^\alpha (1-u_i)^\beta
\prod_{i<j} (u_i-u_j-\theta_i \theta_j)^{-1}
\tag efivetwentynine
$$
and it is identical to a thermal integral
$$
\eqalign{ J_{n+m}(\alpha,\beta,-1) =\prod_0^{n+m-1} {\Gamma(1+\alpha-p+C_p)
\;\Gamma(1+\beta-p+C_p) \over \Gamma(1+\alpha+\beta-(n+m-1)+p-C_p)}\;
& \mu_{n+m}(-1) \cr
  & C_p = \left[{p \over 2}\right] }
\tag efivethirty
$$
Writing the Ansatz (5.5) for $\rho=-1$ we can identify the integers
$M,M'$ as
$$
\eqalign{ M'_p & = C_p \qquad p=0,1,\dots,n-1 \cr
  M_p & = C_{n+p} -n  \qquad p=0,1,\dots,m-1 }
\tag efivethirtyone
$$
There is a certain arbitrariness in this choice.
  We could have taken instead
$M'_p=C_{m+p}-m,\ M_p=C_p$.  This however does not affect the final result.
An argument similar to the one employed in the thermal
 case leads to the same
answer.  Notice the dependence of $M$ on $n$. The same analysis can be
carried out for $J^{\Sigma}_{nm}$ and it leads to
$$
\eqalign{ N'_p & = C_p \qquad N_p = -n + C_{n+p} \cr
  \wtilde{N}'_p & = \wtilde{C}_p \qquad \wtilde{N}_p = -n +
 \wtilde{C}_{n+p} }
\tag efivethirtytwo
$$
Then, when $n\ne 0,\ \wtilde N'_0=-1$ and when $n=0$ (i.e. there is
no $\prod _0^{n-1}$) $\wtilde N_0=-1$.  With this choice the symmetries
$\alpha \wtilde J^{\Sigma}(\alpha,\beta)=\beta \wtilde
J^{\Sigma}(\beta,\alpha)$ and
$J^{\Sigma}(\alpha,\beta)=J^{\Sigma}(\beta,\alpha)$ are automatically
satisfied.  In the reduction from
$J_{n,m}$ to $J_{n-1,m}$ or $J_{n,m-1}$ we have to be careful in taking
into account the $n$-dependence in $M,N$.  The recursion relations
obtained for $\mu_{nm}$ are
$$
\eqalign{ \mu_{nm}(\rho) = (-1)^{1-M_m} e^{-i\pi\rom(m-1)} \rho^n
 {\rho \over 2}
{\gomro \over \Gamma(1+n-\rop m + M_m)} \mu_{n,m-1}(\rho) \cr
\noalign{\hfill (n+m) \hbox{ even} \qquad} }
\tag efivethirtythree
$$
and
$$
\eqalign{ \mu_{nm}(\rho) = (-1)^{1-M_m} e^{-i\pi\rom(m-1)} \rho^n
{\gomro \over \Gamma(1+n-\rop m + M_m)} \mu_{n,m-1}(\rho) \cr
\noalign{\hfill (n+m) \hbox{ odd} \qquad} }
\tag efivethirtyfour
$$
The only difference between these two expressions is the factor of
$\rho/2$.  Iterating the recursion relations we end up in the thermal
case which has already been solved.  After some algebraic manipulations
we arrive at (up to some irrelevant sign), for $n \geq 1$
$$
\eqalign{ \mu_{nm}(\rho) = \rho^{nm} \left({\rho' \over 2} \right)^{M'_n}
& \left(
{\rho \over 2} \right)^{M_m+M'_{n+1}} e^{-i\pi\ropm{n(n-1) \over 4}}
e^{-i\pi\ropm{m(m-1) \over 4}}\cr
  & {\gomrop^n \gomro^m \over \prod_1^{n} \Gamma
(1-\ropp p + M'_p) \prod_1^{m} \Gamma(1+n-\rop p + M_p)} }
\tag efivethirtyfive
$$
For $n=0$ we have the thermal result
$$
\eqalign{ \js_{0,m}(\albr) & = \rho \js_m(\albr) \cr
  \mu_{0,m(}\rho) & = \left( \half \right)^{M'_m} \rho^{M_{m+1}}
{e^{-i\pi\rom{m(m-1) \over 4}} \; \gomro^m \over \prod_1^{n} \Gamma(1-
\rop p + M'_p)} }
\tag efivethirtysix
$$
Finally
$$
\eqalign{ \qquad\jnm(\albr) = & \prod_0^{n-1}
{\Gamma(1+\alpha'+\ropm p+M'_p)
\;\Gamma(1+\beta'+\ropm p+M'_p) \over \Gamma(1+\alpha'+\beta'+\rho'(n-1)
-m-\ropm p-M'_p)} \cr
  & \prod_0^{m-1} {\Gamma(1+\alpha+\rom p+M_p) \;\Gamma(1+\beta+\rom p+
M_p) \over \Gamma(1+\alpha+\beta+\rho(m-1)-
n-\rom p-M_p)}\;\mu_{nm}(\rho)\cr
\noalign{ \hfill $(n+m)$  even \qquad} }
\tag efivethirtyseven a
$$
$$
\eqalign {
  \qquad\jnms(\albr) = \eta & \prod_0^{n-1}
{\Gamma(1+\alpha'+\ropm p+M'_p)\;
\Gamma(1+\beta'+\ropm p+M'_p) \over \Gamma(1+\alpha'+\beta'+\rho'(n-1)
-m-\ropm p-\wtilde{M}'_p)} \cr
  & \prod_0^{m-1} {\Gamma(1+\alpha+\rom p+M_p) \;\Gamma(1+\beta+\rom p+
M_p) \over \Gamma(1+\alpha+\beta+\rho(m-1)-n-\rom p-\wtilde{M}_p)}\;
\mu_{nm}(\rho)\cr
\noalign{ \hfill $(n+m)$ odd \qquad} }
\eqno(efivethirtyseven b)
$$
where
$$
\eqalign{ M'_p & = \left[{p \over 2}\right] \qquad M_p = -n + \left[{n+p
\over 2}\right] \cr
  \wtilde{M}'_p & = \wtilde{C}_p \qquad \wtilde{M}_p = -n + \wtilde{C}_{n+p}
\qquad \hbox{ with } \quad\wtilde{C}_0 = -1,\; \wtilde{C}_{p>0} = \left[
{p \over 2} \right] }
\tag efivethirtyeight
$$
These results can be unified in a way useful for the computation
of $N^{(nm)}_{lk}$, the normalization constants in the conformal blocks.
Define
$$
\hat{J}_{nm}(\albr) = \{ \hbox{same as in \(efivethirtyseven a),
$(n+m)$ even or odd} \}
\tag efivethirtynine
$$
Up to an irrelevant sign one can show that
$$
\eqalign{ \jnm(\albr) & = \hat{J}_{nm}(\albr) \qquad\;(n+m) \hbox{ even} \cr
  \alpha'\tjnms(\albr) & = \eta\hat{J}_{nm}(\albr)\qquad (n+m) \hbox{ odd}}
\tag efiveforty
$$
In terms of ordered integrals,
$$
\eqalign{ \hat{J}_{nm}(\albr) & = \lambda_n(\rho') \epsilon_n(\rho')^{-1}
\lambda_m(\rho) \epsilon_m(\rho)^{-1} \hat{I}_{nm}(\albr) \cr
  & \lambda_m(\rho) = \prod_1^m \;{\siro \over s(\rom)} \qquad
\epsilon_m(\rho) = \prod_0^{m-1} e^{i\pi\rom k} }
\tag efivefortyone
$$
In all our previous results we have systematically ignored some signs
because in the quantities of interest only the square of the normalization
constants is used.
\endpage


\taghead{6.}
\chapter{STRUCTURE CONSTANTS OF THE OPERATOR ALGEBRA}

We have now established all the necessary formulae
 needed for the computation
of the quantities $S^{(m)}_k$ out of which we shall extract the structure
constants.  We find it convenient to deal first with the thermal case.

\section{THERMAL STRUCTURE CONSTANTS}

Recall that we are considering the NS thermal four-point functions,
represented with the help of vertex operators as
$$
\langle \Phi_{\overline{1,t}} (Z_4) \Phi_{1,q}(Z_3) \Phi_{1,n}(Z_2)
\Phi_{1,s}(Z_1) \rangle = \langle V_{\bar\alpha_4}(Z_4) V_{\alpha_3}(Z_3)
V_{\alpha_2}(Z_2) V_{\alpha_1}(Z_1) Q_+^{m-1} \rangle
\tag esixone
$$
It was shown in previous sections  that the four-point correlator takes the
form
$$
\langle V_{\bar\alpha_4} V_{\alpha_3} V_{\alpha_2} V_{\alpha_1}
Q_+^{m-1} \rangle \sim \sum_1^m \skm \left|  \CFkm(\abcr;Z) \right|^2
\tag esixtwo
$$
The quantities $S^{(m)}_k$ were given in \(ethreethirtyonea):
$$
\skm(\abcr) = X_k(\abcr) \left( \nkm(\abcr) \right)^2
\tag esixthree
$$
with
$$
\nkm(\abcr) = (-1)^{m-1} \hat{I}_{m-k}(-1-a-b-c-\rho(m-2),b,\rho)
\;\hat{I}_{k-1}(a,c,\rho)
\tag esixfour
$$
By writing $X^{(m)}_k$ instead of $X_k$ in (6.3) we explicitly indicate
that we make a convenient rescaling of $X_k$ (in other words we choose
a particular value for  $X_m$ in (3.16)).  To compute  $X^{(m)}_k$
we need the matrices $\beta_{kl}(a,b,c,\rho)$ which can be derived from
the Dotsenko and Fateev results [\dotfattwo] through the substitution
$\rho\rightarrow(\rho-1)/2$.  This yields
$$
\eqalign{ \beta_{mk}(\abcr) & =\prod_0^{m-k-1} {s(1+a+b+c+(\rho-1)(m-2)
-\rom i) \over s(b+c+(\rho-1)(m-2)-\rom (m-k-1+i))} \cr
  & \qquad\qquad \prod_0^{k-2} {s(1+b+\rom i) \over
s(b+c+\rom(k-2+i))} \cr
  \beta_{km}(\abcr) & = {\prod_1^{m-1} \siro \over \prod_1^{k-1} \siro
\prod_1^{m-k} \siro} \prod_0^{m-k-1} {s(1+c+\rom(k-1+i)) \over
s(b+c+\rom (m+k-3+i))} \cr
  & \qquad \qquad \prod_0^{k-2} {s(1+b+\rom(m-k+i)) \over
s(b+c+\rom(m-2+i))} }
\tag esixfive
$$
and after some algebra $\ldots$
$$
\eqalign{ X_k^{(m)} & = {\beta_{mm}(a,b)
\beta_{mk}(b,a) \over \beta_{mm}(b,a)
\beta_{km}(a,b) } C^{(m)} \prod_0^{m-2} {s(1+a+\rom i) s(1+c+\rom i) \over
s(a+c+\rom(m-2+i))} \prod_1^{m-1} \siro \cr %
  & = C^{(m)} \prod_1^{k-1} \siro \prod_0^{k-2}
{s(a+\rom i) s(1+c+\rom i) \over s(a+c+\rom(k-2+i)) } \cr
  & \;\prod_1^{m-k} \siro \prod_0^{m-k-1} {s(1+b+\rom i)
s(a+b+c+(\rho-1)(m-2)-\rom i) \over s(a+c+(\rho-1)(m-2)-\rom (m-k-1+i))} }
\tag esixsix
$$
We have introduced the constants $C^{(m)}=(-1)^{m-1}\pi^{2-2m}
\Gamma({1+\rho\over 2})^{2m-2}$ for later convenience.  Repeatedly using
the identity $s(x)\Gamma(x)=\pi/\Gamma(1-x)$ and doing some appropriate
shifts in the arguments of the sine functions in \(esixsix) we obtain
a reasonably nice expression for $S^{(m)}_k$:
$$
\eqalign{ \skm = \left({\rho \over 2}\right)^{2M_{k-1}} & \prod_1^{k-1}
\Delta(\rop i-M_i)
\prod_0^{k-2} \Delta(-a-c-\rho(k-2) + \rom i +M_i) \cr
 & \Delta(1+a + \rom i + M_i) \Delta(1+c+ \rom i +M_i) \cr
\left({\rho \over 2}\right)^{2M_{m-k}} & \prod_1^{m-k} \Delta(\rop i-M_i)
\prod_0^{m-k-1} \Delta(1+a+c+\rho(k-1)+ \rom i + M_i) \cr
  & \Delta(1+b + \rom i + M_i) \Delta(-a-b-c-\rho(m-2)+
\rom i +M_i) }
\tag esixseven
$$
where we have defined $\Delta(x) = \Gamma(x)/\Gamma(1-x)$ and
$M_n = [{n \over 2}]$.\hfil\break
This equation is more symmetrical that it appears at first sight. Since we
restrict ourselves to the thermal subalgebra, the only way to meet the
charge screening requirement is by taking a single conjugate vertex
operator, which we take to be $V_{\overline{\alpha}_4}$.  Defining
$\overline{d}=\alpha_+\overline{\alpha}_4$  and using the charge screening
condition we obtain
$$
\bar d = -1-a-b-c-\rho(m-2)
\tag esixeight
$$
Then the second product of $S^{(m)}_k$ in (6.7) becomes
$$
\eqalign{ \left({\rho \over 2}\right)^{2M_{m-k}} \prod_1^{m-k} \Delta
(\rop i-M_i) &
\prod_0^{m-k-1} \Delta(-b-\bar d-\rho(m-k-1)+\rom i+M_i) \cr
  &  \Delta(1+b+\rom i+M_i) \Delta(1+\bar d+\rom i+M_i) }
\tag esixnine
$$
with the same structure as the first product in \(esixseven).  Introducing
explicitly the Kac labels for the vertex operators, we get the different
parameters :
$$
\eqalign{ a & = \alpha_{1,s} \alp = {1-s \over 2} \rho \cr
  b & = \alpha_{1,q} \alp = {1-q \over 2} \rho }
\qquad\qquad
\eqalign{ c & = \alpha_{1,n} \alp = {1-n \over 2} \rho \cr
  \bar d & = \alpha_{\overline{1,t}} \alp = -1+ {1+\bar t \over 2} \rho }
\tag esixten
$$
where $s,n,q,\bar t$ are positive odd integers (NS sector) related to the
number of screening charges $m-1$ through
$$
m = \half(s+n+q-\bar t)
\tag esixeleven
$$
Then, $S^{(m)}_k$ becomes
$$
\eqalign{ S_k(snq\bar t) = \left({\rho \over 2}
\right)^{2M_{k-1}}&\prod_1^{k-1}
\Delta(\rop i-M_i) \prod_0^{k-2} \quad \Delta(({s+n \over 2}-k+1)\rho +
\rom i+M_i) \cr
  & \Delta(1+{1-s \over 2}\rho+\rom i+M_i) \Delta(1 + {1-n \over
2}\rho+\rom i+M_i) \cr
  \left({\rho \over 2}\right)^{2M_{m-k}} \prod_1^{m-k}\Delta & (\rop i-M_i)
\prod_0^{m-k-1} \Delta(1+({q-\bar{t} \over 2} -m+k)\rho + \rom i + M_i) \cr
  & \quad\Delta(1+{1-q \over 2} \rho +\rom i+M_i) \Delta({1+\bar{t}\over 2}
\rho+\rom i+M_i) }
\tag esixtwelve
$$
{}From this formula we can read off the asymmetric structure constants.  The
index $k$ labels the intermediate channels contributing to the four-point
function.  We can establish also the connection between the number $k$
and the conformal dimension of the field exchanged in the corresponding
internal channel.  This is achieved by choosing a configuration
of superpoints $Z_i$ in (6.1) such that
$$
\left| \Zot \right| \sim \left| Z_{34} \right|
\sim r \ll R \sim \left| Z_{13}
\right| \sim \left| Z_{24} \right|
\tag esixthirteen
$$
Evaluating the four-point function using the OPE of the fields at $Z_1,Z_3$
we can write
$$
\eqalign{ \langle \Phi_{\overline{1,t}} (Z_4) \Phi_{1,q}(Z_3) &
\Phi_{1,n}(Z_2)
\Phi_{1,s}(Z_1) \rangle \sim \cr
  \sim \sum_p & r^{-2(h_s+h_n+h_q+h_{\bar t}-2h_p)} C_{\bar{t}q}^{\bar p}
C_{ns}^p \langle [\Phi_{\overline{1,p}}(Z_3)] [\Phi_{1,p}(Z_1)]
\rangle }
\tag esixfourteen
$$
For brevity we collectively denote by $C^p_{ns}$ the two structure constants
$A^p_{ns}$ and $B^p_{ns}$ and by $[\Phi_{1,p}(Z_1)]$
 the superconformal tower
of descendant fields including in it the factors $Z^{-1/2}_{21}$ when
necessary.  This fine structure of the OPE and of the
 superconformal blocks
is not necessary in the present discussion.  For the choice (6.13) we
obtain
$$
\CFkm(\abcr;Z) = z^{(k-1)(\half+a+c+{\rho \over 2}(k-2))}\;(1+\dots)
\tag esexfifteen
$$
Here again we discard the occasional factors $z^{-1/2}$.  Including the
factors relating (6.1) to (6.2) and after some algebra we arrive at
$$
\eqalign{ p & = s+n+1-2k \cr - \bar p &= -\bar t+q+1-2(m-k+1) }
\tag esixsixteen
$$
Furthermore we can identify the constants
$$
S_k(snq\bar t) = C_{\bar{t}q}^{\bar p} C_{ns}^p
\tag esixseventeen
$$
Finally we can read off the asymmetric structure constants
$$
\eqalign{ C_{ns}^p=\left({\rho \over 2}\right)^{2M_{k-1}} \prod_1^{k-1} &
\Delta(\rop i-M_i) \prod_0^{k-2} \qquad
\Delta(1+{1-s \over 2} \rho + \rom i + M_i) \cr
 & \quad \Delta(1 + {1-n \over 2} \rho + \rom i + M_i) \Delta({1+p \over 2}
\rho + \rom i + M_i) \cr
  C_{\bar{t}q}^{\bar p} = \left({\rho\over 2}\right)^{2M_{l-1}}
\prod_1^{l-1}&
\Delta(\rop i-M_i) \prod_0^{l-2}\qquad \Delta(1+{1-q \over 2} \rho +
\rom i + M_i) \cr
 & \quad\Delta({1+\bar{t} \over 2} \rho + \rom i + M_i) \Delta(1+{1-\bar{p}
\over 2} \rho + \rom i + M_i) }
\tag esixeighteen
$$
where the integers
$$
k=\half(s+n-p+1) \qquad \qquad l=\half(q+\bar p - \bar t +1)
\tag esixnineteen
$$
are now chosen to be functions of the quantum numbers $s,n,p$ and
$q,\overline p,\overline t$. It is very interesting to notice that
although we had to distinguish in our analysis in section 3.1 between
even and odd structure constants, we end up here with a common expression
for both.  This will be the same when we compute the physical structure
constants in the next section.  Using the analyticity properties of
the $\Gamma$ functions, it is easy to see that the structure constants
we have found reproduce the correct fusion rules mentioned in section two.

\section{GENERAL STRUCTURE CONSTANTS}

This subsection follows the steps of the previous one except for the fact
that the computations are more tedious.  We will
 give few details.  Once again
we are interested in
$$
\eqalign{
  \langle \Phi_{\overline{t',t}} (Z_4) \Phi_{q',q}(Z_3) \Phi_{n',n}(Z_2)
\Phi_{s',s}(Z_1) \rangle & = \langle V_{\bar\alpha_4}(Z_4)
 V_{\alpha_3}(Z_3)
V_{\alpha_2}(Z_2) V_{\alpha_1}(Z_1) Q_-^{n-1} Q_+^{m-1} \rangle \cr
  & \sim \sum_{k,l} \slknm \left| \CFlknm(\abcr;Z) \right|^2 }
\tag esixtwenty
$$
The quantity $\slknm$ is given in \(ethreefortyfour)
$$
\slknm(\abcr) = X_l(a',b',c',\rho') X_k(\abcr)
\left( \nlknm(\abcr) \right)^2
\tag esixtwentyone
$$
For $X^{(n)}_l,X^{(m)}_k$ we take the same normalization as in (6.6).  From
\(ethreeforty)  we obtain
$$
\nlknm(\abcr) = \hat{I}_{n-l,m-k}(-a-b-c-\rho(m-2)+n-2,b,\rho) \;
\hat{I}_{l-1,k-1}(a,c,\rho)
\tag esixtwentytwo
$$
Combining $X^{(n)}_l$ with the product of $\Gamma$-functions containing the
primed quantities and similarly for $X^{(m)}_k$ for the unprimed quantities,
we end up after a long calculation with
$$
\eqalign{ \slknm = \rho^{2(l-1)(k-1)} & \left({\rho' \over 2}
\right)^{2 M'_{l-1}} \left({\rho \over 2}\right)^{2 M_{k-1}+ 2M'_{l}} \cr
  \prod_1^{k-1} & \Delta(1-l +\rop i -M_i)
\prod_1^{l-1} \Delta(\ropp i -M'_i) \cr
  \prod_0^{k-2} & \Delta(1+a+\rom i + M_i) \Delta(1+c+\rom i + M_i) \cr
  & \quad \Delta(-a-c-\rho(k-2)+l-1+\rom i + M_i) \cr
  \prod_0^{l-2} & \Delta(1+a'+\ropm i + M'_i) \Delta(1+c'+\ropm i +
M'_i)\cr
  & \quad\Delta(-a'-c'-\rho'(l-2)+k-1+\ropm i + M'_i) \cr
\times \rho^{2(n-l)(m-k)} & \left({\rho' \over 2}\right)^{2 M'_{n-l}}
\left({\rho \over 2}\right)^{2 M_{m-k}+ 2M'_{n-l+1}} \cr
  \prod_1^{m-k} & \Delta(l-n +\rop i-N_i)
\prod_1^{n-l} \Delta(\ropp i -M'_i) \cr
  \prod_0^{m-k-1}\Delta(1+b+ & \rom i + N_i)
\Delta(a+c+\rho(k-1)-l+2+\rom i+
N_i) \cr
  & \qquad\Delta(1-a-b-c-\rho(m-2)+n-2+\rom i + N_i) \cr
  \prod_0^{n-l-1}\Delta(1+b'+ & \ropm i + M'_i)
\Delta(a'+c'+\rho'(l-1)-k+2+\ropm i + M'_i) \cr
  & \qquad\Delta(1-a'-b'-c'-\rho'(n-2)+m-2+\ropm i + M'_i) }
\tag esixtwentythree
$$
where
$$
M_i = 1-l + \left[{l-1+i \over 2} \right] \qquad N_i =  l-n + \left[{n-l+i
\over 2} \right] \qquad M'_i = \left[i \over 2\right]
\tag esixtwentyfour
$$
Using the charge screening condition arising from (6.20) we define
$$
\eqalign{ \bar d & = \bar\alpha_4 \alp = -a-b-c-\rho(m-2) + n-2 \cr
  \bar d' & = \bar\alpha_4 \alm = -a'-b'-c'-\rho'(n-2) + m-2 }
\tag esixtwentyfive
$$
Which helps make the second set of products in (6.23) more similar to the
first one.  Repeating the arguments leading to (6.16) we find the
Kac labels of the intermediate channels
$$
p = s+n+1-2k \qquad\qquad p' = s'+n'+1-2l
\tag esixtwentysix
$$
With the simplifying notation $C^P_{NS}=C^{(p',p)}_{(n',n),(s',s)}$ we
find
$$
\slknm = \wtilde{C}_{\bar T Q}^{\bar P} \cnsp
\tag esixtwentyseven
$$
Finally, introducing the Kac labels for the
parameters $a,b,c\dots$ we obtain
the asymmetric structure constants:
$$
\eqalign{ \cnsp & = \rho^{2(l-1)(l'-1)} \left({\rho' \over 2}
\right)^{2 M'_{l'-1}} \left({\rho \over 2}\right)^{2 M_{l-1}+ 2M'_{l'}}\cr
  &\prod_1^{l-1} \Delta (1-l' +\rop i -M_i)
\prod_1^{l'-1} \Delta(\ropp i -M'_i) \cr
  & \prod_0^{l-2} \Delta ({1+s' \over 2}+ {1-s \over 2}\rho+\rom i+M_i)
\Delta({1+n' \over 2}+ {1-n \over 2}\rho + \rom i + M_i) \cr
  & \qquad \Delta({1-p' \over 2}+ {1+p \over 2}\rho + \rom i + M_i) \cr
  & \prod_0^{l'-2}\Delta ({1+s \over 2}+ {1-s' \over 2}\rho'+\ropm i+M'_i)
\Delta({1+n \over 2}+ {1-n' \over 2}\rho' + \ropm i + M'_i) \cr
  & \qquad \Delta({1-p \over 2}+ {1+p' \over 2}\rho' + \ropm i + M'_i) }
\tag esixtwentyeight
$$
with
$$
\eqalign{ l & = \half(s+n-p+1) \cr l' & = \half(s'+n'-p'+1) } \qquad \qquad
\eqalign{ M_i & = 1-l' + \left[{l'-1+i \over 2} \right] \cr
  M'_i & = \left[i \over 2\right] }
$$
We do not write explicitly the other structure constants
$\wtilde C^{\overline P}_{\overline T Q}$ since they are simply related to
(6.28) by
$$
\wtilde{C}_{\bar{T}Q}^{\bar{P}} = C_{\bar{P}Q}^{\bar{T}}
\tag esixtwentynine
$$
One readily checks that these structure constants ( non-symmetrical )
do not reproduce the correct fusion rules due to some cancellations
between zeroes and poles of various $\Delta$ factors.  The physical
structure constants do agree however with the correct fusion rules, and
they are the subject of the next section.

A direct application of this result is the
evaluation of some surface integrals
corresponding to correlators where the screening charges are integrated over
the whole plane instead of contours.  Consider the three-point function
$$
\eqalign{ J_{l'l}^{(2D)}(a,b,\rho) & = \lim_{R \rightarrow \infty}
R^{4h_M} \langle  V_{\alpha_M}(R,R\eta) V_{\alpha_N}(1,0)
V_{\alpha_S}(0,0) \int \prod_1^{l'} d^2Z'_i\; V_\alm
(Z'_i,\overline{Z}'_i) \cr
  & \phantom{ = \lim_{R \rightarrow \infty}
R^{4h(\alpha_M)} \langle  V_{\alpha_M}(R,R\eta) V_{\alpha_N}(1,0)
V_{\alpha_S}(0,0) \int }\int \prod_1^l d^2Z_i\; V_\alp(Z_i,\overline{Z}_i)
\rangle \cr
  & = \int \prod_1^{l'} d^2Z'_i \int \prod_1^l d^2Z_i \;\xi\;
\prod_{i=1}^{l'} |z'_i|^{2a'} |1-z'_i|^{2b'} \prod_{i<j}^{l'}
|z'_i -z'_j -\theta'_i \theta'_j|^{2\rho'} \cr
  & \qquad\qquad \prod_{i=1}^l |z_i|^{2a} |1-z_i|^{2b}
\prod_{i<j}^l |z_i -z_j -\theta_i \theta_j|^{2\rho} \prod_{i,j}^{l,l'}
|z_i-z'_j-\theta_i \theta'_j|^{-2} \cr}
\tag
$$
where
$$
\eqalign{ \xi & = |1-\alpha_M\alp\sum_i\eta\theta_i-\alpha_M
\alm\sum_i\eta\theta_i'|^2\cr
  a & = \alpha_S \alp \qquad b = \alpha_N \alp \qquad {\rm \; etc.} }
$$
Then with the help of (6.28) we can express it as
$$
\eqalign{ J_{l'l}^{(2D)}(a,b,\rho) = (-1)^{M'_{l'+l}}\; & \pi^{l'+l} l'! l!
\Delta({1-\rho' \over 2})^{l'} \Delta({1-\rho \over 2})^l
\rho^{2l' l} \left({\rho' \over 2}
\right)^{2 M'_{l'}} \left({\rho \over 2}\right)^{2 M_l+ 2M'_{l'+1}} \cr
  \prod_1^l & \Delta(-l' +\rop i -M_i)
\prod_1^{l'} \Delta(\ropp i -M'_i) \cr
  \prod_0^{l-1} & \Delta(1+a+\rom i + M_i) \Delta(1+b+\rom i + M_i) \cr
  & \quad \Delta(-a-b-\rho(l-1)+l'+\rom i + M_i) \cr
  \prod_0^{l'-1} & \Delta(1+a'+\ropm i + M'_i) \Delta(1+b'+\ropm i +
M'_i)\cr
  & \quad\Delta(-a'-b'-\rho'(l'-1)+l+\ropm i + M'_i) }
\tag
$$
with $M_i = -l' +\left[{l'+i \over 2}\right],\; M'_i = \left[i \over
2\right] $.  The proportionality factor in (6.31) is determined by taking
the limit $\rho \rightarrow 0$ for the thermal case, in which limit the
surface integrals decouple and are easily evaluated.  In deriving these
results we choosed the convention that $\int d^2\theta |\theta|^2=1$ and
we omitted to write the factor $|\eta|^2$ that appears in the right-hand
side when $(l'+l)$ is odd.
Equation (6.31) is indeed the generalization to the super case of equation
(B.10) in [\dotfattwo].

\endpage

\taghead{7.}
\chapter{PHYSICAL STRUCTURE CONSTANTS}

All the material necessary for the computation of the physical structure
constants has already been collected in previous sections.  We follow
closely the methodology of [\dotfatthree].  By physical structure
constants we mean the constants $D^P_{SN}$ entering the OPE of
two NS primary fields
$$
\Phi_S(Z_1,\bar Z_1) \Phi_N(Z_2,\bar Z_2) = \sum_P \dsnp \left| \Zot
\right|^{-2(h_S+h_N-h_P)} [\Phi_P(Z_2,\bar Z_2)]_{odd \atop even}
\tag esevenone
$$
For convenience $S=(s',s),\dots$ and no distinction is made between
the odd and even parts of this expansion (this can easily be done by
counting screening charges in the three-point function).  We shall
impose the normalization condition
$$
D_{SS}^1 = 1
\tag eseventwo
$$
corresponding to a diagonal two-point function
$$
\langle \Phi_S(Z_1,\bar Z_1) \Phi_N(Z_2,\bar Z_2) \rangle = \delta_{S,N}
\left| \Zot \right|^{-4h_S}
\tag eseventhree
$$
and to totally symmetric structure constants.  Their determination is
made with the help of the quantities $S^P(TQSN)$ defined in section 6.
Consider a four-point function with the choice of arguments as in
(6.13),
$$
\eqalign{ \langle \Phi_S(Z_4) \Phi_N(Z_3) \Phi_S(Z_2) \Phi_N(Z_1)\rangle
& =\sum_P {\dsnp\dsnp \over r^{4(h_S+h_N-h_P)} } \langle [\Phi_P(Z_3)]
[\Phi_P(Z_1)] \rangle \cr
  \langle \Phi_S(Z_4) \Phi_S(Z_3) \Phi_N(Z_2) \Phi_N(Z_1)\rangle
& =\sum_P{D_{SS}^P D_{NN}^P\over r^{4(h_S+h_N-h_P)}}\langle [\Phi_P(Z_3)]
[\Phi_P(Z_1)] \rangle }
\tag esevenfour
$$
also written in section 6 as
$$
\eqalign{ \langle SNSN \rangle & \sim \sum_P S^P(SNSN) \left| {\cal F}^P(Z)
\right|^2 \cr
  \langle SSNN \rangle & \sim \sum_P S^P(SSNN) \left| {\cal F}^P(Z)
\right|^2 }
\tag esevenfive
$$
With the present normalization the coefficient at the main singularity
corresponding to the identity intermediate channel is equal to unity
whereas $S^1(SSNN)$ generally is not equal to $1$.  Hence the appropriate
definition for the square of the physical structure constants is
$$
\left( \dsnp \right)^2 = {S^P(SNSN) \over S^1(SSNN)}
\tag esevensix
$$
Using the asymmetric structure constants computed in section 6 we arrive at
$$
\left( \dsnp \right)^2 = {\wtilde{C}_{SN}^P\csnp \over \wtilde{C}_{SS}^1
C_{NN}^1} = {C_{PN}^S\csnp \over C_{NN}^1}
\tag esevenseven
$$
We shall shortly prove the relation
$$
C_{SNP} \equiv C_{SN}^{-P} = -{\rho\over 4} \Delta(\rom)\Delta(\ropp)
C_{PP}^1 \csnp
\tag eseveneight
$$
showing that $C^1_{PP}$ plays the role of a metric for raising or lowering
indices.  Substituting in (7.7) we obtain
$$
\dsnp = -4 \rho'\Delta({3-\rho \over 2}) \Delta({1-\rho' \over 2}) \left(
C_{SS}^1 C_{NN}^1 C_{PP}^1 \right)^{-1/2} C_{SNP}
\tag esevennine
$$
As expected we see the symmetry of $D^P_{SN}$ under interchange of pairs of
indices.

The proof of \(eseveneight) is tedious and will be roughly
sketched here. One uses the $\Delta$-functions properties listed below
$$
\eqalign{ \Delta(1-x) & = \Delta(x)^{-1} \cr
  \Delta(x+n) & = (-1)^n \prod_0^{n-1} (i+x)^2 \Delta(x) \cr
  \Delta(x-n) & = (-1)^n \prod_0^{n-1} (-1-i+x)^{-2} \Delta(x) }
\tag eseventen
$$
In the definition of $\csnp$ the bounds of the products are
$$
l' = \half(s'+n'-p'+1) \qquad\qquad l = \half(s+n-p+1)
\tag eseveneleven
$$
whereas in $C_{SNP}=C_{SN}^{-P}$ they are
$$
\wtilde{l}' = \half(s'+n'+p'+1) \qquad\qquad \wtilde{l} =\half(s+n+p+1)
\tag eseventwelwe
$$
We first show that in the ratio $C_{SNP} / \csnp$ the $(s',s)$ and
$(n',n)$ dependence cancel out (up to some power of $\rho$). This is
achieved using the simple relations for the products
$$
\eqalign{ \prod_0^{l-2} f(i) & = \prod_0^{l-2} f(l-2-i) \cr
  \prod_0^{\wtilde{l}-2} f(i) & = \prod_0^{l-2} f(i) \prod_0^{p-1}
f(i+l-1) }
\tag eseventhirteen
$$
Then the $(p',p)$ dependent part of $C_{SNP}$ cancels against $\mu_{l'l}$
(and similarly for the $(p',p)$ dependence in $\csnp$ against
$\wtilde{\mu}_{\tilde{l}'\tilde{l}}$) up to some factors that turn out to
be proportional to $C_{PP}^1$. The NS condition is frequently used
during these simplifications. The determination of the power of $\rho$ in
\(eseveneight) is a rather delicate issue. Actually, $C_{SNP}$ contains
poles and zeroes that cancel each other, and the
 exponent of $\rho$ is a direct
consequence of the regularization procedure.
 We confirmed the consistency of this
procedure by comparing the result obtained for $D_{SN}^{P}$ using either
\(esevenseven) or \(esevennine). More on this ?

\endpage

\taghead{A.}
\appendix
This appendix is devoted to  the computation of the matrix elements
$\beta_{mk}$ entering in the linear expansion
$$
I_m^{(m)}(\abcr;Z) = \sum_{k=1}^m \beta_{mk} \tikm(\abcr;Z)
\tag eaone
$$
{}From the normalization procedure (3.19) for $I^{(m)}_k$ and the relation
between $I^{(m)}_k$ and $\wtilde I^{(m)}_k$ we know
 that when $z\rightarrow 1$,
$\wtilde I^{(m)}_k$ has the singular behavior
$$
\tikm(Z) \sim (1-z)^{(k-1)(1/2+b+c+(k-2)\rho/2)} \; (\hbox{ integral })
\tag eatwo
$$
As in (3.19) the integral is not necessarily a regular function as
$z\rightarrow 1$.  Depending on the values of $k,m$ it may exhibit
a $(1-z)^{-1/2}$ singular behaviour.  Nevertheless the power in front of
the integral in \(eatwo) characterizes the block sufficiently.

The procedure to evaluate $\beta_{mk}$ is to find in  $I^{(m)}_k$ the
same singular behaviour as $z\rightarrow 1$ as in \(eatwo).
  The coefficient
in front of this divergence will only be proportional to
 $\beta_{mk}$ because
we are not taking into account the normalization of the blocks $\tikm(Z)$.
We start with the integral representation of $I^{(m)}_m$
$$
I_m^{(m)} = z^{\Delta_0} \int_0^1 dS_i (1+a_1\eta\theta +a_2 z^{1/2}
\eta\sum\omega_i) \prod_1^{m-1} s_i^a (1-z s_i)^b (1-s_i-{\theta\omega_i
\over z^{1/2}})^c \prod_{i<j} S_{ij}^\rho
\tag eathree
$$
where $\Delta_0=(m-1)({1\over 2} +a+c+{\rho\over 2}(m-2))$.  Performing
on the first $k-1$ variables $s_i$ the change of variables
$$
t_i = {1-z \over 1-s_i+1-z} \qquad\qquad \theta_i = {t_i \over (1-z)^{1/2}}
\; \omega_i
\tag eafour
$$
and letting:
$$
\eqalign{ \epsilon & = t_i(s_i=0) = {1-z \over 2-z} \rightarrow 0
\qquad \hbox{ when } z \rightarrow 1 \cr
  \oeps & = s_{k-1}(t_{k-1}) = 1-(1-z) \; {1-t_{k-1} \over t_{k-1}}
\rightarrow 1 \qquad \hbox{ when } z \rightarrow 1 }
\tag eafive
$$
we obtain after relabelling the $S_i$ and expanding the terms containing
either $\theta$ or $(1-z)^{1/2}$:
$$
\eqalign{ \ikm = & z^{\Delta_0}(1-z)^{(k-1)(1/2+b+c+(k-2)\rho/2)}
\int_\epsilon^1 dT_1 \int_\epsilon^{T_{k-2}} dT_{k-1}
\int_0^\oeps dS_1 \int_0^{S_{m-k-1}} dS_{m-k} \cr
  & (1+a_1\eta\theta +a_2 \zhf(1-z)^{1/2} \eta\sum_1^{k-1}{\theta_i
\over t_i} + a_2\eta\sum_1^{m-k} \omega_i) (1- {c \over (1-z)^{1/2}}
\sum{\theta\theta_i \over 1-t_i}) \cr
  & (1-c \sum{\theta\omega_i \over 1-s_i}) (1-\rho(1-z)^{1/2}\sum
{\theta_i\omega_j \over (1-(1-z){1-t_i \over t_i} -s_j)t_i } ) \cr
  & \qquad\prod_1^{k-1} t_i^{-1-b-c-\rho(k-2)} (1-(1-z){1-t_i \over
t_i})^b (1-t_i)^c \prod_{i<j} T_{ij}^\rho \cr
  & \qquad\prod_1^{m-k} s_i^a (1-z s_i)^b (1-s_i)^c \prod_{i<j}
S_{ij}^\rho \prod_{i,j} (1-(1-z){1-t_i \over t_i}-s_j)^\rho }
\tag easix
$$
As expected from \(ethreetwentyone) we find four
 types of expansions for this
integral depending on the values of $k$ and $m$.  In each case, the
coefficients $\beta'_{mk}$ we are looking for are given by the regular part
of the integral evaluated at $z=1$.  Then the $T_i$ and $S_i$ integrals
decouple and we get
$$
\eqalign{ \iti)\;\beta'_{mk} & = \int_0^1 \prod_1^{k-1} dT_i
t_i^{-1-b-c-\rho(k-2)} (1-t_i)^c \prod_{i<j} T_{ij}^\rho \cr
  & \qquad\int_0^1 \prod_1^{m-k} dS_i s_i^a (1-s_i)^{b+c+
\rho(k-1)} \prod_{i<j} S_{ij}^\rho \cr
  & = I_{k-1}(-1-b-c-\rho(k-2),c,\rho) \;I_{m-k}(a,b+c+\rho(k-1),\rho) }
\tag easeven \iti
$$
$$
\eqalign{ \itii)\;\beta'_{mk}\eta\theta & = -a_2 c\int_0^1 \prod_1^{k-1}
dT_i (\sum{\theta\theta_k \over 1-t_k}) t_i^{-1-b-c-\rho(k-2)} (1-t_i)^c
\prod_{i<j} T_{ij}^\rho \cr
  & \hphantom{=-a_2 c} \int_0^1 \prod_1^{m-k} dS_i (\sum\eta\omega_i)
s_i^a (1-s_i)^{b+c+\rho(k-1)} \prod_{i<j} S_{ij}^\rho \cr
  & = -a_2 c  \wtilde{I}_{k-1}^\Sigma (c,-1-b-c-\rho(k-2),\rho;\theta)
\;I_{m-k}^\Sigma (a,b+c+\rho(k-1),\rho;\eta) }
\eqno(easeven \itii)
$$
$$
\eqalign{ \itiii)\;\beta'_{mk}\eta & = a_2\int_0^1 \prod_1^{k-1}
dT_i t_i^{-1-b-c-\rho(k-2)} (1-t_i)^c \prod_{i<j} T_{ij}^\rho \cr
  & \hphantom{=a_2} \int_0^1 \prod_1^{m-k} dS_i (\sum\eta\omega_i)
s_i^a (1-s_i)^{b+c+\rho(k-1)} \prod_{i<j} S_{ij}^\rho \cr
  & I_{k-1}(-1-b-c-\rho(k-2),c,\rho)
\;I_{m-k}^\Sigma (a,b+c+\rho(k-1),\rho;\eta) }
\eqno(easeven \itiii)
$$
$$
\eqalign{ \itiv)\;\beta'_{mk}\theta & = -c\int_0^1 \prod_1^{k-1}
dT_i (\sum{\theta\theta_k \over 1-t_k}) t_i^{-1-b-c-\rho(k-2)} (1-t_i)^c
\prod_{i<j} T_{ij}^\rho \cr
  & \hphantom{=-c} \int_0^1 \prod_1^{m-k} dS_i
s_i^a (1-s_i)^{b+c+\rho(k-1)} \prod_{i<j} S_{ij}^\rho \cr
  & = -c\wtilde{I}_{k-1}^\Sigma (c,-1-b-c-\rho(k-2),\rho;\theta)
\;I_{m-k}(a,b+c+\rho(k-1),\rho) }
\eqno(easeven \itiv)
$$
Using the relation with the $\hat I_n$ integrals:
$$
\eqalign{ \alpha \wtilde{I}_n^\Sigma(\albr;\eta) & = \eta \hat{I}_n(\beta,
\alpha,\rho) \cr
  (1+\alpha+\beta+\rho(n-1)) & \;I_n^\Sigma(\albr;\eta) = \eta \hat{I}
(\albr) }
\tag eaeight
$$
together with
$$
a_2 = 1+a+b+c+\rho(m-2) = 1+a+b+c+\rho(k-1) +\rho(m-k-1)
\tag eanine
$$
we reduce \(easeven) to a single expression
$$
\beta'_{mk} = \epsilon_1 \hat{I}_{m-k}(a,b+c+\rho(k-1),\rho) \;
\hat{I}_{k-1}(c,-1-b-c-\rho(k-2),\rho)
\tag eaten
$$
with $\epsilon_1=1$ except for case \itiv) where $\epsilon_1=-1$.

In order to obtain the coefficients $\beta_{mk}$ we only need to divide
$\beta '_{mk}$ by the normalization factor of the superconformal block
$\wtilde I^{(m)}_k$,
$$
\beta_{mk}(\abcr) = {\beta'_{mk}(\abcr) \over \wtilde{N}_k^{(m)}(\abcr)}
\tag eaeleven
$$
The normalization factor $\wtilde N^{(m)}_k$ is derived from
$ N^{(m)}_k$ by using \(ethreeseventeen).  A careful analysis of the four
possible cases leads to
$$
\widetilde{N}_k^{(m)}(\abcr) = \epsilon_2 N_k^{(m)}(b,a,c,\rho)
\tag eatwelve
$$
with $\epsilon_2=1$ except for case \itiii) where $\epsilon_2=-1$.
  The last
relation together with \(ethreetwentysix) enables us to write
$$
\beta_{mk} = {\hat{I}_{m-k}(a,b+c+\rho(k-1),\rho) \;
\hat{I}_{k-1}(c,-1-b-c-\rho(k-2),\rho) \over
\hat{I}_{m-k}(-1-a-b-c-\rho(m-2),a,\rho)
\;\hat{I}_{k-1}(b,c,\rho) }
\tag eathirteen
$$
($\epsilon_1$ and $\epsilon_2$ combine and cancel the sign $(-1)^{m-1}$
entering $N^{(m)}_k$).  The $\mu$ factors  appearing in the $\hat I$
integrals cancel out and we are left with a ratio of products of
$\Gamma$-functions.  Using $\Gamma(x) \Gamma(1-x)=\pi/s(x)$ repeatedly
we finally obtain
$$
\eqalign{ \beta_{mk} = (-1)^{m-1}\prod_0^{m-k-1} & {s(1+a+b+c+
(\rho-1)(m-2)-\rom i) \over s(b+c+(\rho-1)(m-2)-\rom (m-k-1+i))} \cr
  & \prod_0^{k-2} {s(1+b+\rom i) \over s(b+c+\rom(k-2+i))} }
\tag eafourteen
$$
This result is exactly the same as the result obtained by Dotsenko and
Fateev [\dotfattwo] (3.16) provided we implement in their
formulae the substitution $\rho\rightarrow (\rho-1)/2$.
  The sign difference
$(-1)^{m-1}$ arises from the difference between our definition of
conformal blocks and that of [\dotfattwo].  This result was expected, as
explained in the text, but it provides a rather non-trivial test of
our evaluation of the normalization integrals, the main computational
difficulty in this paper.

\endpage
\refout
\endpage

\centerline{\bf Figure captions }

Fig.3.1. Contour ordering for $\jkm(Z)$.

Fig.3.2. Contours for the analytic continuation of $\jkm$.

Fig.3.3. Contour ordering for $\tjkm(Z)$.

Fig.3.4. Contour ordering for $\jlknm(Z)$.

Fig.4.1. Ordering of contours chosen in (4.1).

Fig.4.2. Contour ordering in (4.2,3).

Fig.4.3. Explicit definition of contour ordering.

Fig.4.4. Contours used in the definition of  $J^\Sigma(p,0,n-p)$.

Fig.4.5. Pulling one contour in $J(0,2m)$.

Fig.5.1. Integration contours in (5.1).

Fig.5.2. Contours used for the evaluation of the even integral
$J_{p'+q',p+q}$.

Fig.5.3. Contours used for the evaluation of the odd integral
$J^{k'}_{nm}$.
\endpage

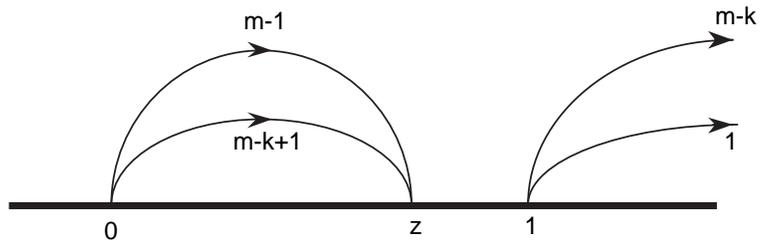
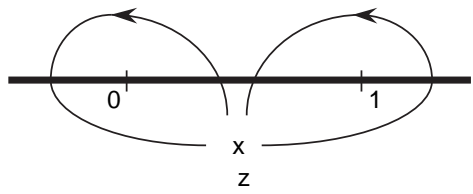
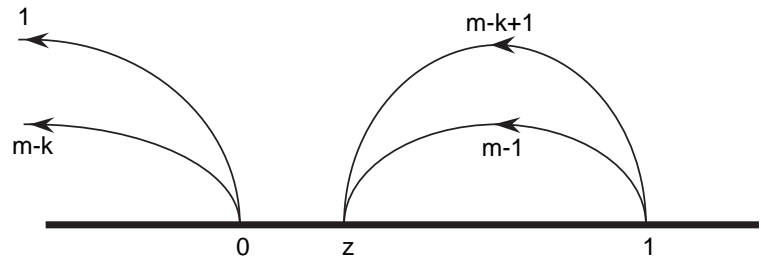
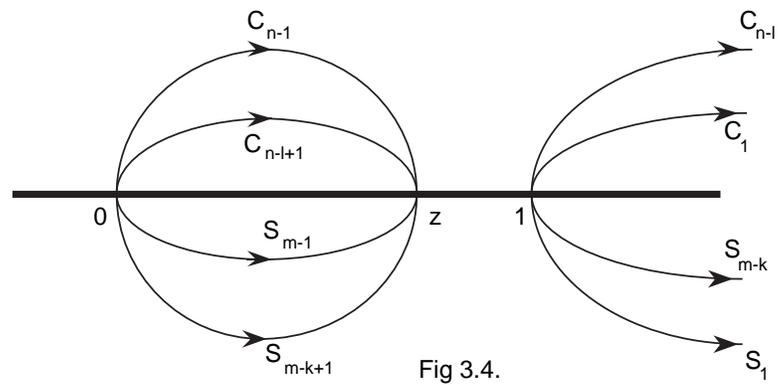
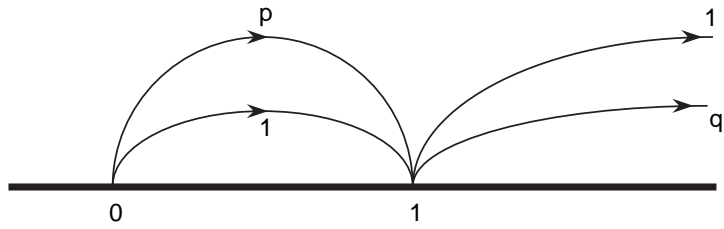
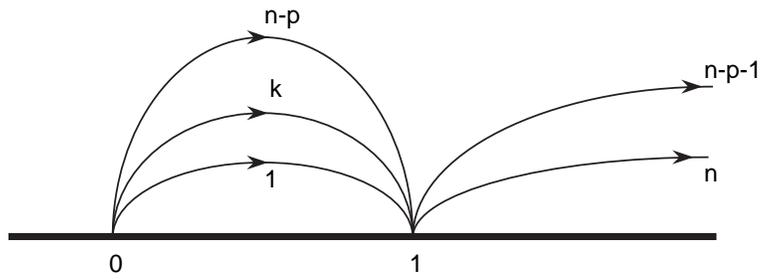
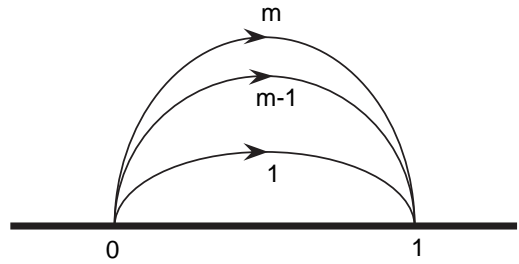
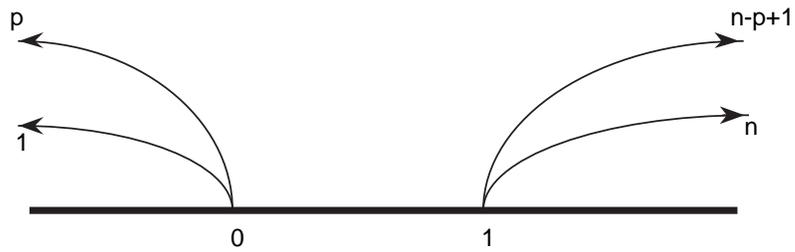
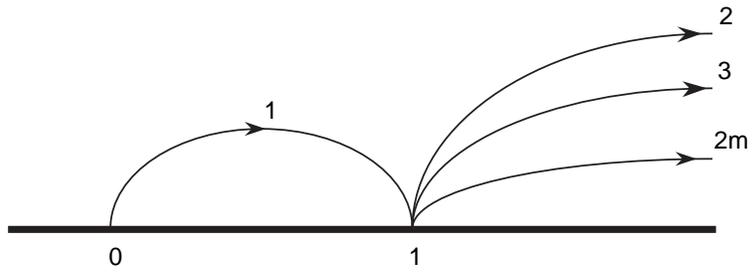
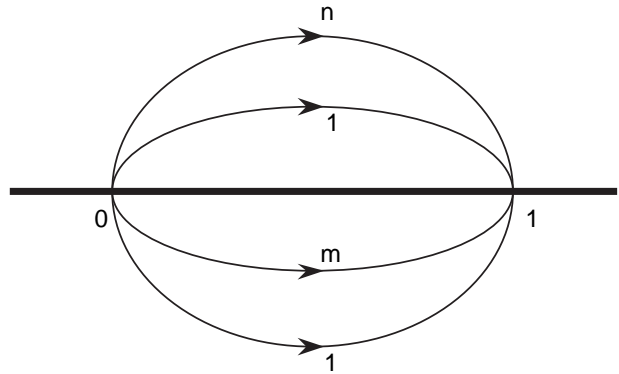
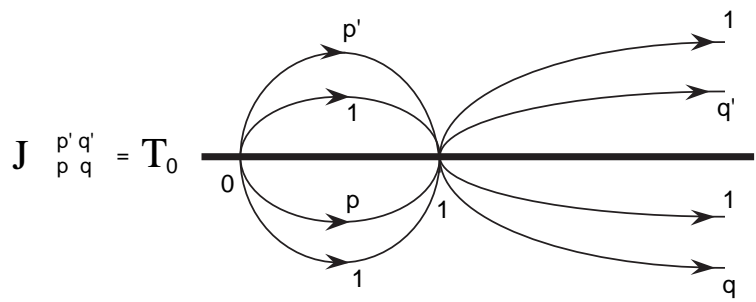
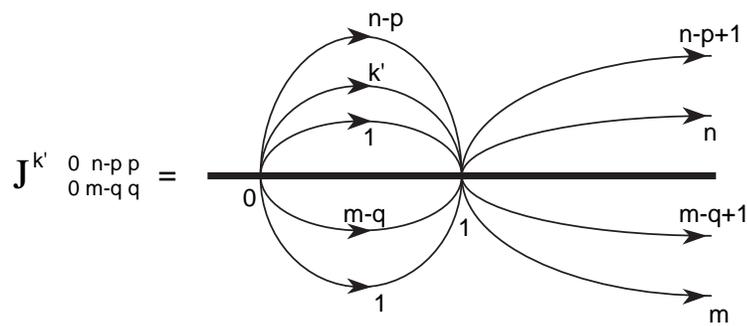
\end